\documentclass[%
 reprint,
 nofootinbib,
 amsmath,amssymb,
 aps,
]{revtex4-2}

\usepackage{graphicx}% Include figure files
\usepackage{dcolumn}% Align table columns on decimal point
\usepackage{bm}% bold math
\usepackage{comment}
\usepackage{xcolor}
\usepackage{soul}
\usepackage{ulem}
\providecommand{\lP}{\left(}
\providecommand{\rP}{\right)}
\providecommand{\lC}{\left[}
\providecommand{\rC}{\right]}
\providecommand{\lCB}{\left\{}

\providecommand{\rp}{\right.}

\providecommand{\PP}[1]{ \left( {#1} \right) }

\providecommand{\bb}[1]{ \left| {#1} \right| }
\providecommand{\av}[1]{ \left\langle {#1} \right\rangle}

\begin{document}

\preprint{APS/123-QED}

\title{Polar order, shear banding, and clustering in confined active matter}
\homepage{\label{ESI}Electronic Supplementary Information (ESI) available.}

\author{Daniel Canavello}\thanks{These authors contributed equally to the work}
\author{Rubens H. Damascena}\thanks{These authors contributed equally to the work}
\author{Leonardo R. E. Cabral}\email{leonardo.cabral@ufpe.br}
\author{Clécio C. de Souza Silva}\email{clecio.cssilva@ufpe.br}
\affiliation{
Departamento de Física, Centro de Ciências Exatas e da Natureza, Universidade Federal de Pernambuco, Recife--PE, 50670-901, Brasil
}

\date{\today}

\begin{abstract}
We investigate the collective behavior of sterically interacting self-propelled particles confined in a harmonic potential. Our theoretical and numerical study unveils the emergence of distinctive collective polar organizations, revealing how different levels of interparticle torques and noise influence the system. The observed phases include the shear-banded vortex, where the system self organizes in two concentric bands rotating in opposite directions around the potential center; the uniform vortex, where the two bands merge into a close packed configurations rotating uniformly as a quasi-rigid body; and the orbiting polar state, characterized by parallel orientation vectors and the cluster revolving around the potential center, without rotation, as a rigid body. Intriguingly, at lower filling fractions, the vortex and polar phases merge into a single phase where the trapped cluster breaks into smaller polarized clusters, each one orbiting the potential center as a rigid body. 

\end{abstract}

\maketitle

\section{Introduction} \label{sec.intro}

Active matter encompasses a wide range of biological and synthetic entities, including swimming bacteria, microalgae, living tissue, and artificial swimmers, capable of converting energy absorbed from the environment into a propelling force that drives them far from equilibrium~\cite{Marchetti2013review,Ramaswamy2010,Bechinger2016,Fodor2018}. 
As a result of their motility and mutual interactions, these systems exhibit a wealth of self-organized collective phenomena, such as flocking, motility-induced phase separation, dynamical clustering, and collective self-optimization~\cite{Szabo2006,Fily_2012,reichhardt2014absorbing,Giavazzi2018,Pince2016,Geyer_2019,Deblais2018,A.C.Quillen}. Many of these phenomena stem from the ability of self-propelled particles to reorient their propulsion direction when interacting with each other. The reported mechanisms behind this polar alignment extends from ad hoc aligning rules, such as the Vicsek interaction, to simple steric forces and hydrodynamic interactions~\cite{Fily_2012,Giavazzi2018,martin-gomez2018,Caprini2020,Sansa2021}. In bulk, aligning interactions result in different kinds of orientational orders, like ferromagnetic, nematic and vortex. These studies have significantly contributed to our understanding of the underlying principles driving self-organization and pattern formation in active matter.

In many situations of interest, active matter interact with confining walls, obstacles, or potential energy landscapes, produced by optical and acoustic tweezers, nonuniform hydrodynamic flows, and phoretic fields~\cite{pototsky2012active,brotto2013hydrodynamics,hardouin2019reconfigurable,Takatori2016,schmidt2021non,hardouin2022active}. Confinement plays a crucial role in shaping the behavior and properties of active matter systems~\cite{Wioland_2016,PhysRevLett.110.268102,C7SM00999B,PhysRevE.107.024606,PhysRevResearch.5.023196,2020PNAS.11711901L} and can induce novel collective phenomena that differ from their bulk counterparts, such as the fluid pump phase predicted for confined particles interacting hydrodynamically~\cite{Hennes_2014}. 
Moreover, confinement is an invaluable tool to investigate non-equilibrium properties of active particles, specially those that distinguish them from passive Brownian systems, as for instance noise-induced escape problems that violate detailed balance~\cite{Woillez2019,Wexler2020,Rubens2023}. However, the impact of confinement on the collective dynamics and phase behavior of self-propelled polar particles remains relatively unexplored. 

The influence of confinement extends beyond forces to include torques that realign the orientation of active particles. These aligning torques are often overlooked in models of confined active matter, despite their significant physical consequences. For instance, in the case of a single confined particle, the interplay of motility and the confining force and torque gives rise to a dynamical phase transition where the particle develops orbital motion within the confining region~\cite{Dauchot2019,Rubens2022,Rubens2023}. This phenomenon has been observed in diverse systems, from geometrically confined motile bacteria~\cite{codutti2022} and microalgae~\cite{Ostapenko2018} to a minirobot on a parabolic reflector~\cite{Dauchot2019}. Similarly, active Janus colloids have been shown to experience an aligning torque induced by a phoretic gradient ~\cite{Liebchen2019}. Despite recent progress in understanding the properties arising from aligning torques at the level of a single active particle, their correlation with collective phenomena in confined active matter remains unclear.

\begin{figure*}[tb]
\includegraphics[width= \linewidth]{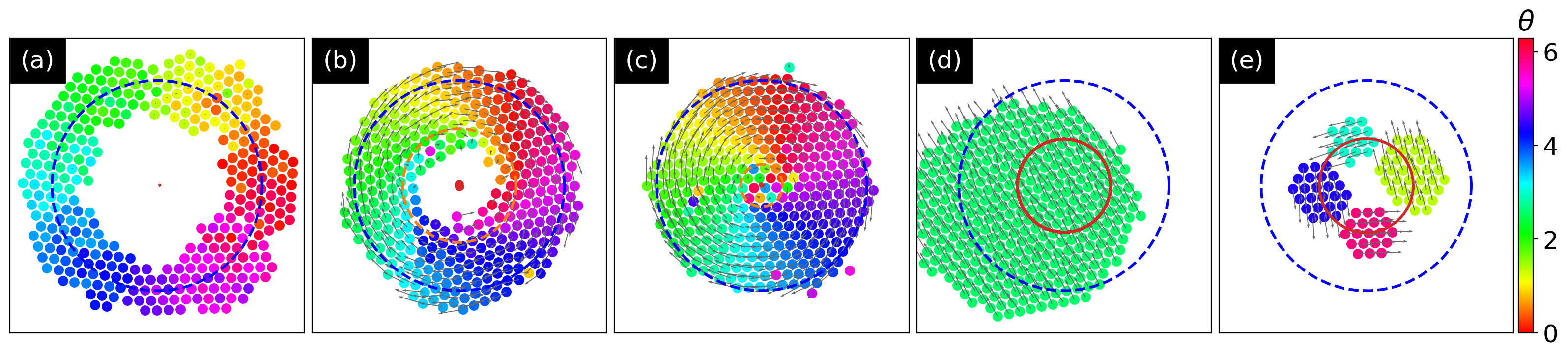}
\caption{Snapshots of positions (dots), orientations (colors) and velocities (arrows) representative of the main phases observed in the system: 
(a) radially polarized, (b) shear-banded vortex, (c) uniform vortex, (d) orbiting ``ferromagnetic'' crystal, and (e) multi-cluster. 
In all cases, noise is turned off and the particle diameter is $\sigma=0.1 R_\text{CI}$, where $R_\text{CI}$ is the radius of the critical isocline (blue dashes) of the harmonic confinement. For (a) to (d) $N=350$ and the values of $\beta$ are, respectively, $0.5, 1.85$, $5.0$ and $5.0$, while for (e) $N=93$ and $\beta = 5.0$. The solid red lines depict the trajectory of the centroid of each cluster. The dashed orange line in (b) and (c) denotes the theoretical boundary between the shear bands (see text).
}
\label{fig:PhasesColor}
\end{figure*}

In this work, we investigate collective phenomena in a model system of monodisperse self-propelled particles trapped by a harmonic potential and interacting with each other via simple steric forces and torques. The contribution of the steric torques and the torque induced by the confining force is regulated by the angular mobility $\beta$. The higher $\beta$, the more relevant is the role played by torques in the the overall dynamics. 
Fig.~\ref{fig:PhasesColor} 
shows snapshots of positions, orientations (color coding) and velocities (arrows) of the particles for different values of noise intensity $D$, angular relaxation rate $\beta$, and particle size $\sigma$ representative of some of the different phases observed in our study (see also the Videos in the Electronic Supplementary Information, ESI). For small $\beta$ and at low noise, the active particles are observed to arrange themselves as a radially polarized fluid, with ring-shaped distribution (a). In increasing $\beta$, the system rotates without revolving as either a shear-banded vortex (b), for moderate filling fractions, or a uniform vortex (c), for large filling fractions. At higher values of $\beta$, the system revolves, without rotation, as a highly polarized (``ferromagnetic'') crystal (d). Finally, at low filling fractions, the system self-organizes in multiple polarized clusters orbiting the potential (e). 

This paper is organized as follows. In Section~\ref{sec:model}, we present the model and derive analytical results for the orbiting polar and vortex phases in the deterministic (quasi-)rigid-body limit. In Section~\ref{sec:numerics}, we present detailed numerical results on all observed phases and their dependence on the control parameters. In particular, we validate and discuss the limitations of our theoretical modelling on the deterministic vortex and orbiting polar phases. We also discuss possible mechanisms for the observation of the shear-banding transition, which is not captured by our model, and the effect of noise on all observed phases. The conclusions are presented in Section~\ref{sec:conclusions}.

\section{Model and analytical results} \label{sec:model}

\subsection{Equations of motion} \label{sec:eom}

We consider a 2D system of $N$ active Brownian particles (ABP), at positions $\bm{r}_i = (x,y)$ and with orientation vectors $\bm{n}_i = (\sin \theta, \cos \theta)$, confined in an isotropic harmonic potential, $V(\bm{r}_i)=\frac{1}{2}\kappa\bm{r}_i^2$. Their dynamics are modeled by overdamped Langevin equations of motion,
\begin{align}
    \label{eq:motion}
    \dot{\bm{r}}_i &= v_0 \bm{n}_i + \mu\bm{F}_i + \sqrt{2D_t}\bm{\xi}_i(t), \\
    \dot{\theta}_i &= \beta (\bm{n}_i \times {\bm{F}}_i)\cdot \hat{\bm{z}} + \sqrt{2D}\zeta_i(t),
    \label{eq:orientation-part}
\end{align}
where $v_0$ denotes the propulsion speed, which is the same for all particles, $\hat{\bm{z}}$ is the unit vector perpendicular to the $xy$ plane, $\mu$ ($\beta$) is the translational (angular) mobility, and $\bm{F}_i = -  \kappa\bm{r}_i - \nabla_i\sum_{j\neq i}^{N} U(|\bm{r}_i-\bm{r}_j|)$ sum up the conservative forces acting on particle $i$. The steric interactions are modeled by a truncated Lennard-Jones (WCA) potential:  $U(r)=4\epsilon[(\sigma/r)^{12}-(\sigma/r)^6]-\epsilon$, for $r<2^{1/6}\sigma$, and $U(r)=0$, for $r>2^{1/6}\sigma$, where $\sigma$ is the particle size and $\epsilon$ is the interaction energy scale. In this work we fix $\epsilon=0.1 v_0^2/(\mu^2\kappa)$. The translational, $\bm{\xi}_i(t)$, and rotational, $\zeta_i(t)$, noise terms satisfy $\langle \bm{\xi}_i(t)\cdot\bm{\xi}_j(t') \rangle = 2\delta_{ij}\delta(t - t')$ and $\langle \zeta_i(t) \zeta_j(t') \rangle = \delta_{ij}\delta(t - t')$, respectively, with $D_t$ and $D$ representing the respective diffusion constants. Also, %{In this work} 
we focus mainly on the so-called active noise regime, which is dominated by rotational diffusion and thus we set $D_t = 0$ ~\cite{Howse_2007,volpe2011,Drescher2011}. The effect of $D_t$ is discussed briefly in the Electronic Supplementary Information (ESI).

The first term in the right-hand side of Eq.~\eqref{eq:orientation-part} represents the restoring torque produced by the total force $\bm{F}_i$ acting on particle $i$. Since from Eq.~\eqref{eq:motion} $\bm{n}_i\times\mu\bm{F}_i=\bm{n}_i\times\dot{\bm{r}}_i$, a non-zero restoring torque appears every time $\bm{n}_i$ and $\dot{\bm{r}}_i$ misalign. This term adds considerable complexity to the collective system dynamics as the terminal orientation of each particle results from a compromise between the orientations of the neighboring particles and that of the local confining force, which are all constantly changing as the system evolves in time. The role of the restoring torque to the overall dynamics depends crucially on the parameter $\beta$, which is a measure of how fast the particle's orientation relaxes towards the direction of the local force field. For small $\beta$, the relaxation time grows making it difficult for the particle to reach the preferred orientation. In this case, the restoring torque plays a negligible role and one restores the conventional ABP model.

\subsection{Length scales and filling fraction}
\label{sec.scales}

The deterministic, one-particle limit of equations~\eqref{eq:motion} and \eqref{eq:orientation-part} presents two steady state solutions for radially symmetric confinement~\cite{Dauchot2019}: (i) The climbing phase, where the particle climbs the potential well until reaching the force balance condition $V'(R_\text{CI}) = v_0/\mu$; and (ii) the orbiting phase, where the particle describes a circular orbit of radius $R$ satisfying $R=v_0/\beta V'(R)$~\cite{Rubens2023}. For the harmonic potential, $V(r)=\frac{1}{2}\kappa r^2$, these length scales read:
\begin{equation}
    R_\text{CI} = \frac{v_0}{\mu\kappa}, ~
    R = \sqrt{\frac{v_0}{\kappa\beta}}
\end{equation}
The transition from climbing to orbiting takes place when $\beta$ becomes larger than the critical value $\beta_c=\mu^2\kappa/v_0$.

The force balance condition establishes a critical isocline (CI) across which the particle can never cross from inside out for any $\beta$. When many particles are present, the force balance condition also involves interparticle interactions. In this case, particles can cross the critical isocline assisted by other ones. This is particularly the case when one exceeds the maximum number of particles fitting the circle of radius $R_\text{CI}$. Therefore, it is convenient to define the filling fraction as the ratio 
\begin{align}
    f=N\frac{A_\text{SPP}}{A_\text{CI}} = \frac{N\sigma_h^2}{4R_\text{CI}^2}
    \label{eq:FillingFraction}
\end{align} 
where $A_\text{SP}=\pi\sigma_h^2/4$ is the effective area occupied by one particle, with $\sigma_h\simeq2^{1/6}\sigma$ being the interparticle distance in a head-on collision, and $A_\text{CI}=\pi R_\text{CI}^2$ is the area enclosed by the critical isocline. We have observed that for $f\gtrsim 1$ the particles typically organize themselves in a single cluster, while for small $f$ the system can break into multiple clusters or in a mixture of clusters and isolated particles.

From here on, we shall adopt a unit system where length is expressed in units of $R_\text{CI}=v_0/(\mu\kappa)$, time in units of $t_0=1/(\mu\kappa)$, and energy in units of $v_0^2/(\mu^2\kappa)$.

\begin{figure*}[htb!]
%\centering
\includegraphics[width= \linewidth]{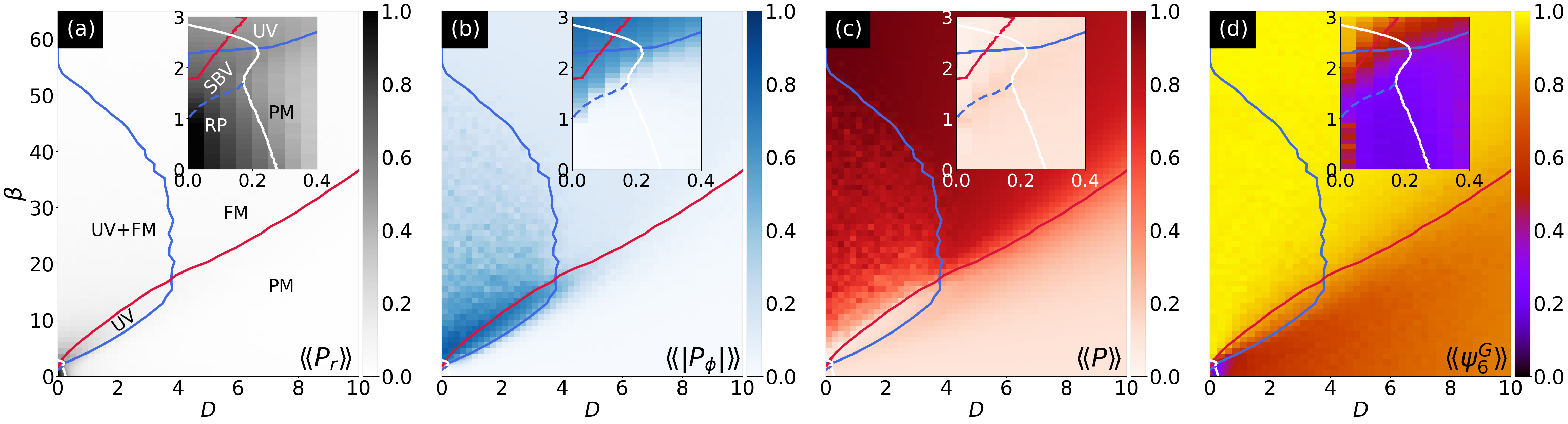}
\caption{
Colormaps of the time and random noise averages of the (a) radial, $\av{\av{P_r}}$, (b) azimuthal, $\av{\av{P_\phi}}$, and (c) modulus of the total, $\av{\av{\bb{\bm{P}}}}$, polarizations, as well as (d) the hexatic order parameter, $\av{\av{\psi^G_6}}  $, as function of ($\beta$, $D$) for $N = 93$ and $\sigma = 0.2$.
The red, blue, and white lines represent, respectively, the boundaries at which the ferromagnetic (FM), uniform vortex (UV), and paramagnetic (PM) phases become unstable. The blue dashed line separates the shear-banded vortex (SBV) and the radially polarized (RP) phases. 
}
\label{fig:PhaseDiagram}
\end{figure*}

\subsection{Deterministic steady-state solutions of close-packed clusters} \label{sec:analytical}

Here, we analyse the collective motion of the system when the particles move as close-packed clusters in the absence of noise. In this situation, the dimensionless version of Eqs.~\eqref{eq:motion} and~\eqref{eq:orientation-part} become
\begin{align}
    & \dot{\bm{r}}_k +  \bm{r}_k - \sum_{\substack{l \neq k}}^N {\bm{f}_{kl}} = \bm{n}_k,
  \label{eq:motion_deterministic}\\
  & \dot{\theta}_k = \beta (\bm{n}_k\times \dot{\bm{r}}_k)\cdot \hat z, % = -\beta \CC{\bm{n}_k\times \PP{\bm{r}_k - \sum_{\substack{l \neq k}}^N {\bm{f}_{kl}}}}\cdot \hat z,
    \label{eq:orientation_deterministic}
\end{align}
where \(\bm{f}_{kl} = f(r_{kl}) {\bm{r}_{kl}}/{r_{kl}}\) are arbitrary pair-wise forces (in which the WCA model yields a particular case), $\bm{r}_k = \PP{r_k \cos\phi_k,\,r_k \sin\phi_k}$, $\bm{n}_k = \PP{\cos\theta_k,\,\sin\theta_k}$, $\bm{r}_{kl} = \bm{r}_k - \bm{r}_l$, and $r_{kl}  = |\bm{r}_k - \bm{r}_l|$. 

By summing Eq.~\eqref{eq:motion_deterministic} for all the particles, the internal forces $\bm{f}_{kl}$ cancel out and we have an equation of motion for the system centroid $\bm{R} = \sum_{k=1}^N \bm{r}_k / N$,
\begin{align}
  & \dot{\bm{R}} + \bm{R} = \dfrac{1}{N}\sum_{k=1}^N \bm{n}_k = \bm{P},
  \label{eq:centroid-motion}
\end{align}
where we have also defined the total polarization $\bm{P}$ of the cluster.

It is possible to show (see the ESI for details) that if all particles have the same orientation, that is $\bm{n}_k=\bm{P}$ for all $k$, the cluster moves as a rigid body. In this case, the dynamics of the polar cluster is fully described by Eq.~\eqref{eq:centroid-motion}, supplemented by $\dot{\theta}_k=\beta(\bm{P}\times\dot{\bm{R}})\cdot\hat{\bm{z}}$ for all $k$, and thereby is equivalent to that of a single particle in a harmonic confinement~\cite{Rubens2022,Dauchot_2019}. For $\beta > 1$, the steady state solution of this problem is an orbital motion of the cluster, with centroid position $R$, angular velocity $\Omega$, and orientation angle $\theta_{\rm c}$ given by,
\begin{align}
R & = \beta^{-1/2}, \label{eq:polarcluster_R}\\
\Omega & = \pm \sqrt{\beta - 1}, \label{eq:polarcluster_Omega}\\ 
%\chi_{\rm c} & = \theta_{\rm c} - \varphi = \arccos\PP{\beta^{-1/2}} \quad \Rightarrow\quad 
\theta_{\rm c}(t) & = \arccos\PP{\beta^{-1/2}} \pm\sqrt{\beta - 1} t  + \theta_{\rm c}\PP{0}.\label{eq:polarcluster_theta}
\end{align} 

Another case of interest is the one in which we have a cluster of particles rotating around the origin. In such vortex phase, $\bm{P} = 0$, since $\bm{R} = 0$. A relation between the particles angular velocities $\dot{\phi}_k$ and their radial positions $r_k$ can be obtained from the vector product between $\bm{r}_k$ and Eq.~\eqref{eq:motion_deterministic}. After summing all the particles contributions, internal torques cancel, and we find
\begin{align}
\sum_k r_k^2 \dot{\phi}_k =  \sum_k r_k \sin{\chi}_k,
\label{eq:angvel_vortex}
\end{align} 
where we define the tilt angle $\chi_k = \theta_k - \phi_k$ with respect to the radial position.  As shown in the ESI, a strict rigid body rotation with angular velocity $\omega$ would require $\dot{\theta}_k = \omega$, which is admissible only for particles within the region $r>1/\beta$. Instead, we model the system still assuming that all particles rotate with the same angular velocity and keep their radial positions, but relaxing the $\dot{\chi}=0$ condition. 
Adopting this procedure, which we shall refer to as the quasi-rigid-body (QRB) model, Eq.~\eqref{eq:orientation_deterministic} yields (see ESI for more details)
\begin{align}
\dot{\chi}_k = \PP{\beta r_k \cos\chi_k - 1}\omega.
\label{eq:vortex_chi-diff_eq}
\end{align}
The solution of this equation reveals two distinct regimes of motion depending on the particle distance to the origin. For $r_k > 1/\beta$, $\chi_k$ depends monotonically on time, 
\begin{align}
\chi_k = 2\arctan\lC\lP 
 \dfrac{C_+ + C_- e^{-t/\tau_\omega}}{C_+ - C_- e^{-t/\tau_\omega}}\rP
\sqrt{\dfrac{\beta r_k-1}{\beta r_k+1}}\rC, \label{eq:chi_out}
\end{align}
where \(C_\pm = \tan\lP\chi_k(0)/2\rP \pm \sqrt{\lP \beta r_k-1\rP/\lP \beta r_k+1\rP}\), $\chi_k(0)$ is the initial value of $\chi_k$ at $t = 0$ and \(\tau_\omega = 1/\omega\sqrt{\beta^2 r_k^2 - 1}\). Notice that for long times, i.e., $t \gg \tau_\omega$, the asymptotic value \(\chi_k = 2\arctan\sqrt{\lP \beta r_k-1\rP/\lP \beta r_k+1\rP}\) is achieved. This gives  \(\cos \chi_k = 1/\beta r_k\), which is the result expected by assuming $\omega = \dot{\theta}_k$.
Meanwhile, for $r_k < 1/\beta$, $\chi_k$ has oscillatory response given by 
\begin{align}
\chi_k = -2\arctan\lC 
\sqrt{\dfrac{1-\beta r_k}{1+\beta r_k}} \tan\lP
\nu t/2 - C_0
%\sqrt{1-\beta^2 r_k^2}\dfrac{\omega t}{2} - C_0
\rP
\rC, \label{eq:chi_in}
\end{align}
where $\nu = \sqrt{1-\beta^2 r_k^2}\omega$  
is the angular frequency of $\chi_k$ 
and $C_0 = \arctan\lC \sqrt{\lP 1-\beta s_k\rP/\lP 1+\beta s_k\rP} \tan\lP\chi_k(0)/2\rP\rC$.

In order to know the radial and angular components of $\bm{n}_k$, namely $\sin\chi_k$ and $\sin\chi_k$ respectively, we computed their time averages from the above time dependencies of $\chi_k$. For $r > 1/\beta$ the time averages for $t \gg t_\omega$ are given by the asymptotic behavior of $\chi_k$. On the other hand, for $r < 1/\beta$, we calculate the time averages, $\av{\sin\chi_k}$ and $\av{\cos\chi_k}$ over one period of oscillation, $T = 2\pi/\nu$. Therefore,
\begin{align}
    \av{\sin\chi_k} & = \lCB 
    \begin{array}{l l}
    \sqrt{1 - \dfrac{1}{\beta^2 r_k^2}}, & \text{ for } r_k > 1/\beta, \\
    0, & \text{ for } r_k < 1/\beta,
    \end{array}
    \rp, \label{eq:vortex_sinchi}\\
    \av{\cos\chi_k} & = \lCB
    \begin{array}{l l}
    \dfrac{1}{\beta r_k}, & \text{ for } r_k > 1/\beta, \\
    \dfrac{
1 - \sqrt{1 - \beta^2 r_k^2}
}{
\beta r_k}, & \text{ for } r_k < 1/\beta.
    \end{array}
    \rp\label{eq:vortex_coschi}
\end{align}
Now, we return to Eq.~\eqref{eq:angvel_vortex} and compute its time average, substituting  the appropriate dependencies of $\sin\chi_k$ given by Eq.~\eqref{eq:vortex_sinchi}. Assuming $\av{\dot{\phi}} \approx \av{\omega}$ we find the time-average angular velocity of the rotating cluster 
\begin{align}
    \av{\omega} =  \dfrac{\displaystyle\sum_k{}^{\prime}\, \sqrt{
r_k^2 - \dfrac{1}{\beta^2}
}}{\displaystyle\sum_k r_k^2 },
\label{eq:omega_vortex}
\end{align}
where $\displaystyle\sum_k{}^{\prime}$ means sum over only the particles outside of $r = 1/\beta$, while $\displaystyle\sum_k$ is the sum over all the particles.
Moreover, substituting Eq.~\eqref{eq:vortex_coschi} in Eq.~\eqref{eq:orientation_deterministic} enable us to obtain an approximate time average of $\dot{\theta}_k$, i.e., 
 \begin{align}
 \!\!\!\av{\dot{\theta}_k} =
 \lCB 
 \begin{array}{rr}
     \av{\omega}, & \text{ for } r_k > \beta^{-1}, \\
    \PP{1 - \sqrt{1 - \beta^2 r_k^2}}\av{\omega}, & \text{ for } r_k < \beta^{-1}. 
 \end{array}
 \rp
     \label{eq:dot-theta}
 \end{align}

It is important to mention that these results for the vortex phase implies nonzero $\dot{\bm{r}}_k$ in  Eq.~\eqref{eq:motion_deterministic}, particularly for $r < 1/\beta$. These lead to nonzero radial distances variations $\delta r_k$ of the particles positions.  Therefore, our model for the vortex phase is expected to hold for $\delta r_k \ll r_k$. 

\section{Numerical results} 
\label{sec:numerics}

To investigate in further detail the different dynamical phases and phase transitions exhibited by this system, we numerically integrated  Eqs.~\eqref{eq:motion} and \eqref{eq:orientation-part}, using a second-order stochastic Runge-Kutta algorithm, exploring a wide range of values of angular mobility $\beta$, rotational noise intensity $D$, and filling fraction $f$. 
In Fig.~\ref{fig:PhaseDiagram}, we present the dynamical phase diagram in the $\beta D$ parameter plane for a system of $N=93$ self-propelled particles of size $\sigma=0.2$, which corresponds to an intermediate-valued filling fraction $f=1.17$, in the active noise ($D_t=0$) regime. The heatmaps correspond to time and random noise averages, $\langle\!\langle\,\cdots\rangle\!\rangle$,\footnote{
 Unless stated otherwise, time averages were computed over $100 t_0$ after a $300 t_0$ waiting time while noise averages were performed over 50 independent realizations of the stochastic forces.%, each one starting from a random initial configuration of the particles positions and orientations.
 \label{Footnote}} 
of the total, radial, and azimuthal polarizations, given respectively by
\begin{align}
    \bm{P} = \sum_{i = 1}^{N}\frac{\bm{n}_i}{N}, 
    ~
    {P}_r = \sum_{i = 1}^{N}\frac{\bm{n}_i\cdot\hat{\bm{r}}_i}{N}, 
    ~
    {P}_\phi = \sum_{i = 1}^{N}\frac{\bm{n}_i\cdot\hat{\bm{\phi}}_i}{N}, 
    \label{eq:PolarOP}
\end{align}
where $\hat{\bm{r}}_i= \bm{r}_i/|\bm{r}_i|$ and $\hat{\bm{\phi}}_i=\hat{\bm{z}}\times\hat{\bm{r}}_i$, and the hexatic order parameter, defined as
\begin{equation}
    \psi_6^G = \left|\frac{1}{N}\sum_{i=1}^N\left[
    \frac{1}{{\cal N}_i} \sum_{j=1}^{{\cal N}_i} e^{i6\theta_{ij}}\right]\right|,
    \label{eq:psi6}
\end{equation}
where $\mathcal{N}_i$ is the number of nearest neighbors of particle $i$ and $\theta_{ij}$ is the angle of the link between $i$ and neighbor $j$ with respect to the $x$ axis.

These quantities allow for identifying in the parameter space the distinct phases represented by the configurations shown in Fig.~\ref{fig:PhasesColor} (a)-(d). The radially polarized (RP) state is characterized by large (close to one) $\langle\!\langle P_r\rangle\!\rangle$ and vanishingly small $\langle\!\langle P_\phi\rangle\!\rangle$ and $\langle\!\langle|\bm{P}|\rangle\!\rangle$. 
The vortex phases present two distinct behaviors: 
%, as we shall discuss in more detail below: 
for $\beta\gtrsim1$, $\langle\!\langle P_r\rangle\!\rangle$ is still large, but $\langle\!\langle P_\phi\rangle\!\rangle$ is non-zero. The small azimuthal alignment of the particle orientations is just enough to make the system rotate in either clockwise or anticlockwise direction as a shear-banded vortex (SBV); for $\beta\gg1$ or large values of $f$, the system rotates uniformly and typically has $\langle\!\langle P_\phi\rangle\!\rangle>0.5$ and $\langle\!\langle P_r\rangle\!\rangle<0.5$, which reflects the global azimuthal alignment of the particle orientations. This uniform vortex (UV) phase is also characterized by large hexactic order as a result of the close-packed arrangement of the particles. The orbiting  ferromagnetic (FM) phase is characterized by a high degree of orientational and translational order as revealed by the large values of $\langle\!\langle|\bm{P}|\rangle\!\rangle$ and $\langle\!\langle|\psi_6^{G}|\rangle\!\rangle$. It dominates a large part of the parameter space and coexist with the vortex phases for small $D$ and intermediate $\beta$ values. Finally, in the small $\beta$ and large $D$ region all polar order parameters vanish, which characterize the unpolarized or paramagnetic (PM) state. In the following subsections, we discuss in more detail the properties of each polar phase and the effect of changing $\beta$, $D$, number of particles, and particle size. 

Fig.~\ref{fig:PhaseDiagram} also features estimates of the stability boundaries of the phases, shown as continuous white, red, and blue lines. They were obtained by initializing the system in a given phase and sweeping $\beta$ and $D$ until the main polar order parameter characterizing that phase falls below a threshold value (we choose 0.5 as the threshold for all polar order parameters, but see below for details). Beyond this point, the phase is considered to be extinct and is replaced by a new phase, either in a continuous or abrupt way. This allows a reasonably accurate identification of the stability limit of the phases for the cases of abrupt transition, such as the UV to FM and FM to UV transitions, and a rough estimate for smooth transitions, such as all transitions involving the paramagnetic state. The consistency of these criteria was validated by videos of the system evolution at points nearby the estimated boundaries. The transition between the radially polarized and the shear-banded vortex state requires a different criterion. In that case, since rotation commences as soon as a small azimuthal polarization appears, we use the more stringent condition $\langle\!\langle P_\phi\rangle\!\rangle>0.1$ (dashed blue line) as an estimate of the break down of the RP phase in favour of the SBV phase. The SBV phase has typically a larger value of $\langle\!\langle P_r\rangle\!\rangle$ as compared to the UV phase, so the $\langle\!\langle P_\phi\rangle\!\rangle=0.5$ line can be taken as a rough estimate for the SBV to UV transition.

We have also analysed the impact of non-negligible  translational noise on the observed phases and their respective boundaries. For that, we considered the case where the random noise is dominated by thermal fluctuations, for which $D_t=\sigma^2D/3$~\cite{Howse_2007,volpe2011}. %The main effect of $D_t$ is to displace the phase boundaries to lower values of D (see Supplemental Material \dcma{ESI}). No new phases were detected. \dcma{This sentence is a bit confusing}
While no new phases emerged in this case, notable changes were observed in some phase boundaries compared to the $D_t=0$ scenario depicted in Fig.\ref{fig:PhaseDiagram}. The impact is particularly pronounced for large values of $\beta$ and $D$, leading to significant alterations in the FM-PM phase boundary. Conversely, transitions between the RP phase and the vortex phases as well as the RP-PM phase boundary, occurring at small $D$ and relatively small $\beta$, remain essentially unchanged (see Supplemental Material, {ESI}).

\subsection{Radially polarized phase} \label{sec:RadialPolarizedPhase}

The radially polarized phase is reminiscent of the climbing phase, where non-interacting active particles under radial confinement are predicted to accumulate at a certain distance away from the potential center~\cite{pototsky2012active,Fily2014,Takatori2016,Malakar2020,Caraglio2022,Nakul2023}. This distance coincides with the radius of the critical isocline for small $D$, as discussed in Sec.~\ref{sec.scales}. Here, because of the steric interparticle interactions, this accumulation manifests itself in different ways. For $D=0$ and $\beta\lesssim1$, the particles exhibit a close-packed configuration in a ring-shaped crystal [see Fig.~\ref{fig:PhasesColor}-(a) for a representative configuration], resulting in high values of the hexatic order parameter. At non-zero but small values of $D$, the hexatic order decays considerably and the system exhibits a fluidlike behavior, while still keeping the radial polar order (see Supplementary Video 1, ESI). Upon increasing $D$ further at small $\beta$, the system loses the radial polar order and enters the unpolarized phase. On the other hand, increasing $\beta$ at $D=0$ makes the system to transition to the vortex phase and subsequently, at a higher value of $\beta$, to the orbiting polarized state as explained below. 

\subsection{The ferromagnetic and uniform vortex phases: global properties and phase coexistence}

\begin{figure}[tb!]
%\centering
\includegraphics[width= \linewidth]{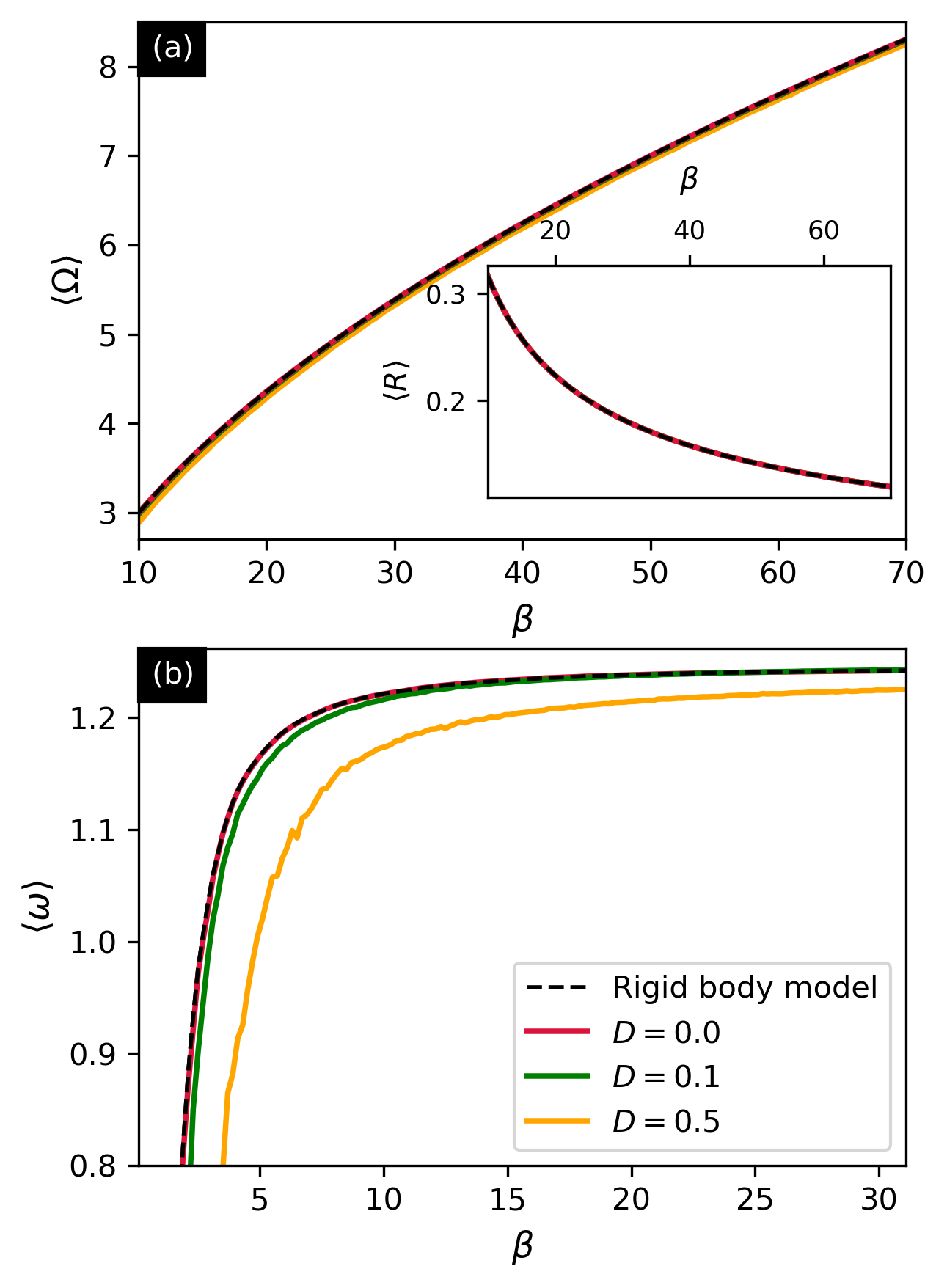}
\caption{
Comparison between the analytical predictions and numerical results for a cluster with $N = 93$ and $\sigma = 0.2$ in (a) the orbiting polar and (b) the vortex phases. The solid lines represent the numerical results of the mean angular frequency (a) and (b) and centroid radial distance (inset) for different values of $D$. The dashed lines correspond to the analytical results [Eqs.~\eqref{eq:polarcluster_Omega}, \eqref{eq:polarcluster_R}, and \eqref{eq:omega_vortex}]
}
\label{fig:FrequencyComparation}
\end{figure}

In the orbiting ferromagnetic (FM) phase [Fig.~\ref{fig:PhasesColor}-(d)], the particles form a single close-packed cluster that revolves around the potential center but never rotates around its own centroid. Additionally, all orientation vectors are  parallel to each other and rotate at the same rate as the angular speed of the centroid (see Supplementary Video 2, ESI). These attributes suggest the FM phase can be well described by our rigid-body model presented in Sec.~\ref{sec:analytical}, which reduces the problem to that of a single particle with constant orbital radius and angular speed given by Eqs.~\eqref{eq:polarcluster_R} and \eqref{eq:polarcluster_Omega}, respectively. In Fig~\ref{fig:FrequencyComparation}-(a), 
we confront these analytical results with measurements of the mean angular speed and radial position of the centroid of a system of $N=93$ particles of diameter $\sigma=0.2$ in the orbiting polar phase. The simulations were performed as a function of $\beta$ for different $D$. The numerical results are shown to be in excellent agreement with the theory, even in the presence of moderate noise. Indeed, as already anticipated in the diagrams of Fig.~\ref{fig:PhaseDiagram}, the orbiting polar phase is highly resilient to noise, being stable even at high values of $D$, specially for high angular mobility. %The main effect of noise is to induce fluidlike motion of particles at the cluster edge, without significant changes in the hexactic order parameter 

The uniform vortex (UV) phase is also characterized by a close-packed arrangement of the particles, but now the system rotates around its own centroid which remains  essentially fixed at the potential minimum. In Fig.~\ref{fig:FrequencyComparation}-(b), we show the average angular speed of the particles along $100 t_0$ time window as a function of $\beta$ for different values of $D$ and compare with the analytical formula Eq.~\eqref{eq:omega_vortex}. For $D=0$, the simulation results are in excellent agreement with the model. However, the agreement becomes considerably poor for $D>0.1$. Notice that $D=0.5$ is well within the stability range of the UV phase, which becomes unstable only at $D\gtrsim4$ for $10<\beta<30$. Still, the mean angular speed of the system at this noise level is considerably smaller than the $D=0$ limit, suggesting that the UV phase is typically much more sensitive to fluctuations than the FM phase.

\begin{figure}[htb!]
%\centering
\includegraphics[width= \linewidth]{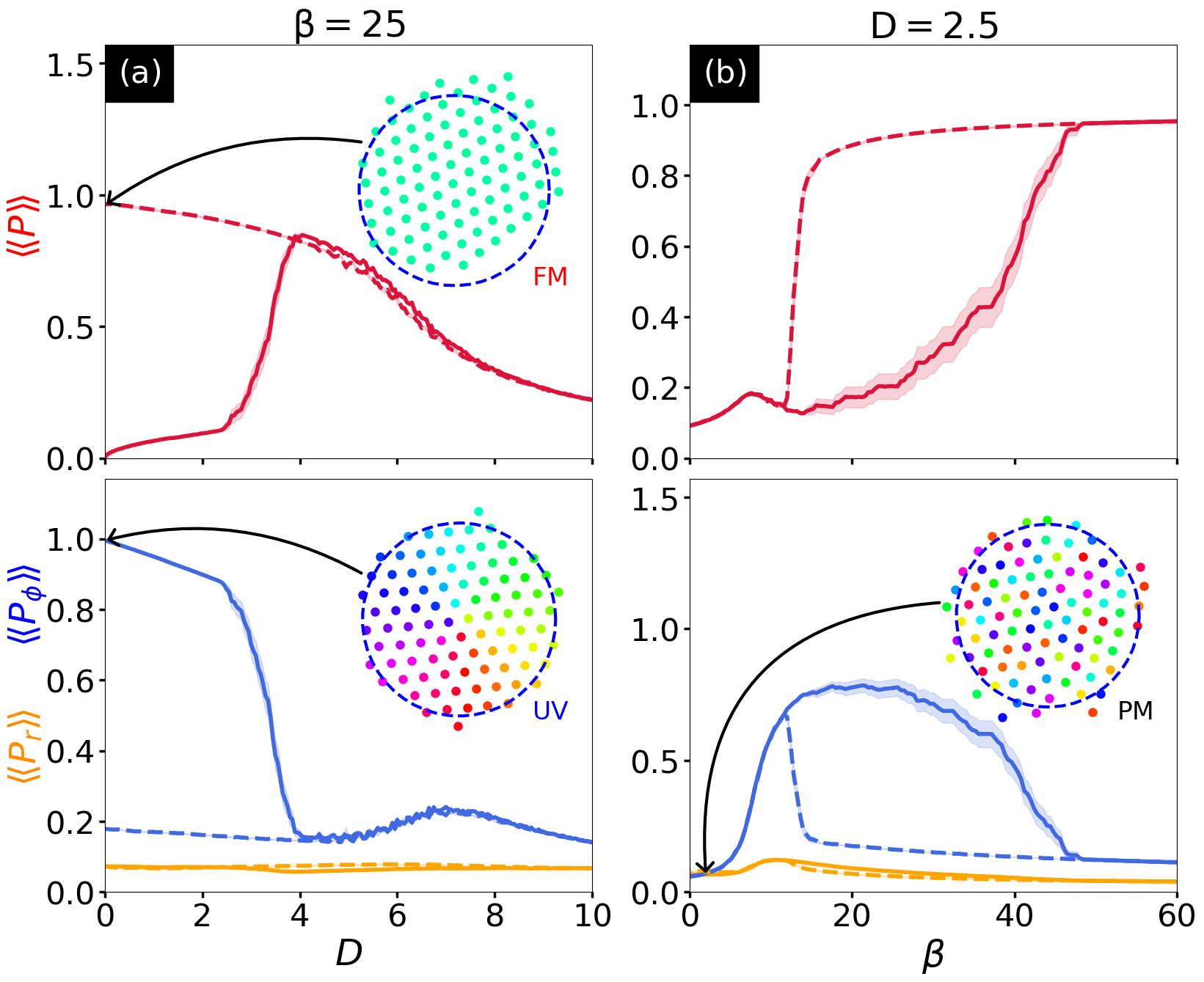}
\caption{
(a) Time and noise averages of the polar order parameters, $\langle\!\langle|\bm{P}|\rangle\!\rangle$ and $\langle\!\langle|P_\phi|\rangle\!\rangle$, calculated for fixed $\beta=25$ and sweeping $D$ from zero up to 10 (solid lines) and back to zero (dashes). The shadows correspond to standard deviations from the mean. (b) Same as (a), but fixing $D=2.5$ and sweeping $\beta$ from 0 to 60 and back. The simulations were performed for $N=93$ and $\sigma=0.2$ as in Fig.~\ref{fig:PhaseDiagram}. The insets depict representative snapshots of the system at the points indicated by the arrow. The color scheme is the same as in Fig.~\ref{fig:PhasesColor}.
}
\label{fig:SweepsOnPhaseDiagram}
\end{figure}

The FM and UV phases have a large intersection where both phases coexist, that is, depending on the initial conditions, the system self-organizes in either FM or UV phases, with the FM (UV) being more probable for higher (lower) values of $\beta$, as indicated by the color gradients in Fig.~\ref{fig:PhaseDiagram} (b) and (c). As a result, the transition from UV to FM and back exhibits pronounced hysteretic behavior as illustrated in Fig.~\ref{fig:SweepsOnPhaseDiagram}. 
In panel (a) the system is initialized in the vortex phase at $\beta = 25$ and $D = 0$. Upon increasing $D$ slowly up to 10, the vortex phase decays into the orbiting polar phase, as indicated by the sharp drop in the azimuthal polar order parameter accompanied by an increase in the total polarization. Then, the FM phase continuously crosses over to the PM phase. Now, decreasing $D$ from 10, the transition from PM to FM follows approximately the same path, thus emphasizing the continuous character of this transition. In contrast, by decreasing $D$ even further, the FM phase never decays back to the vortex phase, following a different path from the upward branch. Similar hysteretic phenomena is also observed by sweeping $\beta$ up and down at fixed $D=2.5$, as shown in panel (b). In this case, the system is initialized in the unpolarized state which is the only one available for $(\beta,D)=(0,2.5)$. In increasing $\beta$, it crosses over reversibly to the uniform vortex state and at $\beta\simeq40$ decays to the FM phase. Reversing $\beta$ from 60 down to 0, the FM phase decays back to UV phase only at $\beta\simeq13$, thus revealing a pronounced hysteresis.

\begin{figure}[t!]
%\centering
\includegraphics[width= \linewidth]{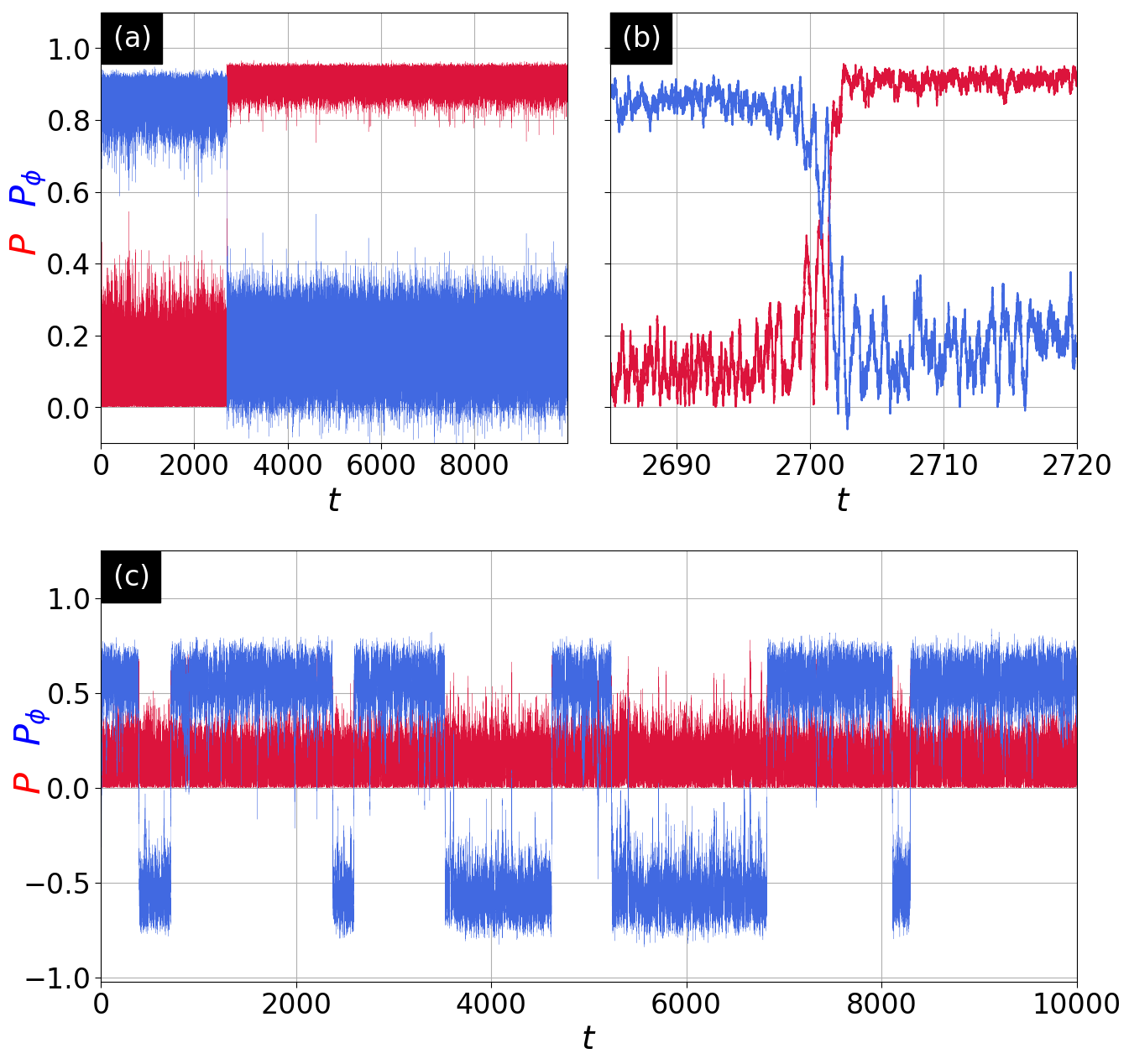}
\caption{
Time series of the total polarization modulus (red) and the azimuthal polarization (blue) featuring: (a) and (b) a transition between UV and FM phases for $\beta = 25$ and $D = 2.85$ [(b) shows a zoomed-in view of the transition]; (c) multiple switchings between clockwise and anticlockwise vortex phases for $\beta = 9.2$ and $D = 2.5$.  }
\label{fig:TemporalSeriesExample}
\end{figure}

Fluctuations can often induce spontaneous transitions  between coexisting states. To check the possibility of similar noise-induced switching between distinct phases in our system, we selected a few $(\beta,D)$  points and, for each of them, calculated the azimuthal and total polarization order parameters as a function of time along a $10^4t_0$ time window, starting from one of the coexisting phases.  
In Fig.~\ref{fig:TemporalSeriesExample}%
we illustrate two kinds of noise-induced switching between coexisting dynamical states observed in this study. Panel (a) shows a typical phase switching from the uniform vortex to the ferromagnetic phase at a point within the UV-FM coexisting zone, $(\beta,D)=(25,2.85)$. After initializing the system at the uniform vortex state, a sudden drop of the azimuthal order parameter and concomitant rise of the total polarization take place at $t\simeq2700t_0$. Conversely, by initializing the system in the FM state at the same point of the parameter space, no transition is observed within the simulated time window. This suggests that at this point the FM state is more stable than the UV state. In panel (c), we show the time evolution of the system for a point in the phase diagram where only the UV phase is stable. The data reveal multiple switchings of the sign of the azimuthal order parameter, thus reflecting transitions between clockwise and counterclockwise UV rotation. Unlike the UV-FM spontaneous transitions, here the inversions of the UV rotation to one direction or the other occur at similar rates, as a result of the symmetry of both states.

\subsection{Anatomy of the uniform vortex phase}
\label{sec:UVanatomy}

\begin{figure}[tb!]
%\centering
\includegraphics[width= \linewidth]{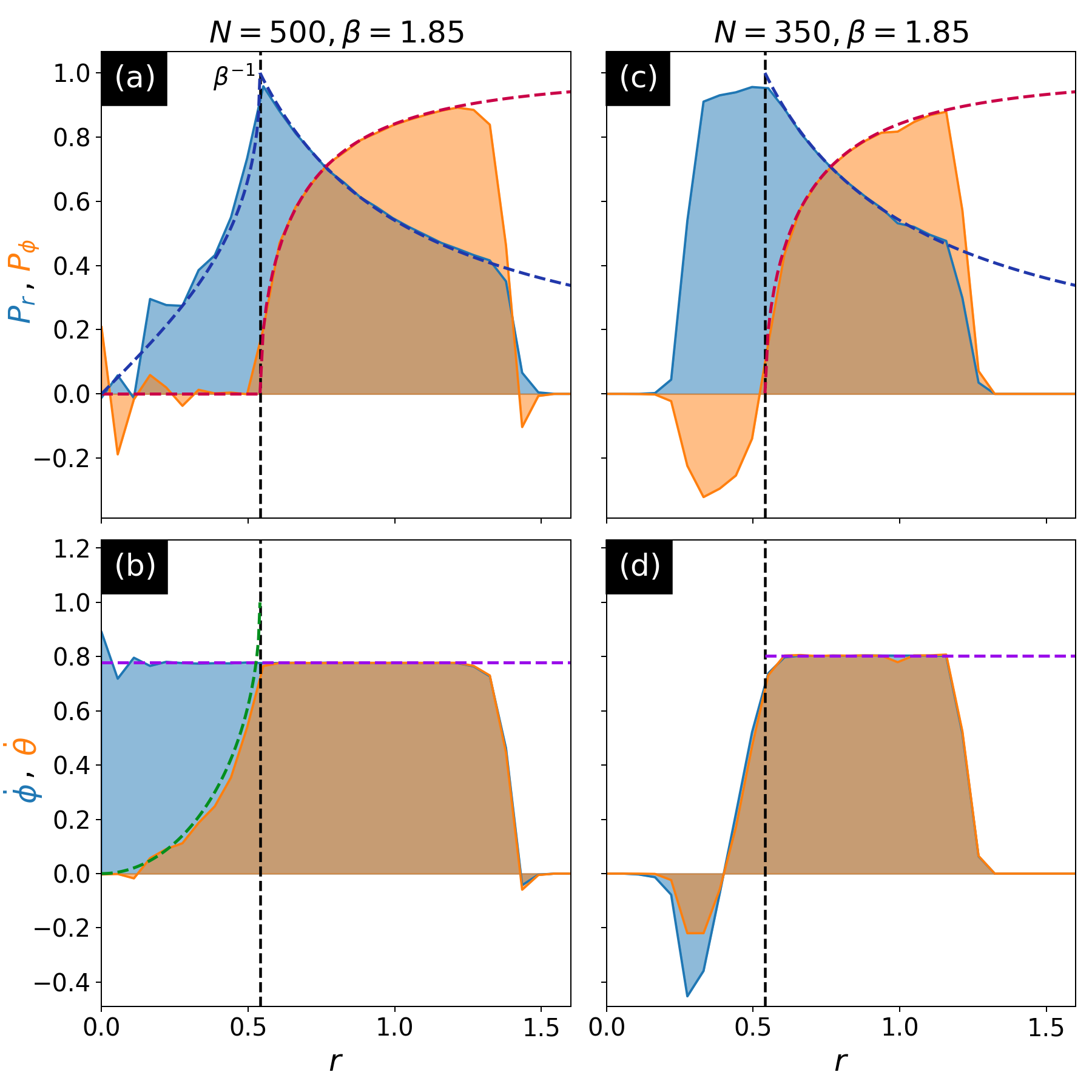}
\caption{Time-averaged radial distribution of $P_R$, $P_\phi$, $\dot \theta$ and $\dot \phi$ calculated from the simulations (filled areas) and predicted by the model (dashed lines). The red and blue dashed lines are given by Eqs.~\eqref{eq:vortex_sinchi} and \eqref{eq:vortex_coschi}, while the purple and green lines were obtained from Eqs.~\eqref{eq:omega_vortex} and \eqref{eq:dot-theta}. The situation in (a) and (b)  ($N=500$ and $\beta = 1.85$) corresponds to the {UV} state, which is fully captured by the model. For (c) and (d) ($N=350$ and $\beta = 1.85$), the system is in the SBV state, in which case the model fails for $r < \beta^{-1}$ (see text).
}
\label{fig:ByDistance}
\end{figure}

As mentioned above, the uniform vortex phase is characterized by uniform rotation of the particles, that is, all $\dot{\phi}_k$ are essentially the same. In contrast, the polarization vectors of the particle showcase a highly non-uniform radial distribution. To illustrate and better understand these intriguing local properties, we consider a smaller particle size, $\sigma=0.1$. This allows for investigating clusters with considerably larger number of particles, which in turn renders better local statistics while keeping the filling fraction close to one. In Fig.~\ref{fig:ByDistance}
(a) and (b) we show the time-averaged radial distributions of $P_r$, $P_\phi$, $\dot{\phi}$, and $\dot{\theta}$ for $N=500$ particles and $(\beta,D)=(1.85,0)$. For these parameters, the particles form a close-packed cluster that rotates uniformly, that is $\dot{\phi}(r)\simeq\text{constant}$. Nevertheless, all other measured quantities present strikingly different behavior in two different regions. For $r>1/\beta$, these quantities follow closely the theoretical prediction for a perfect rigid-body vortex: $P_r = \cos\chi = 1/\beta r$ and $P_\phi = \sin\chi = \sqrt{1 - 1/\beta^2r^2}$. For $r<1/\beta$, the rigid-body condition $\dot{\theta}=\dot{\phi}$ breaks down. However, since $\dot{\phi}$ is still uniform, the approximations that lead to Eq.~\eqref{eq:vortex_chi-diff_eq} still apply. Indeed, the profiles of $P_r$, $P_\phi$, and $\dot\theta$ in this region are well described by Eqs.~\eqref{eq:vortex_sinchi}, ~\eqref{eq:vortex_coschi}, and ~\eqref{eq:dot-theta}, respectively, in the full range of $r$.

\begin{figure}[t]
%\centering
\includegraphics[width= \linewidth]{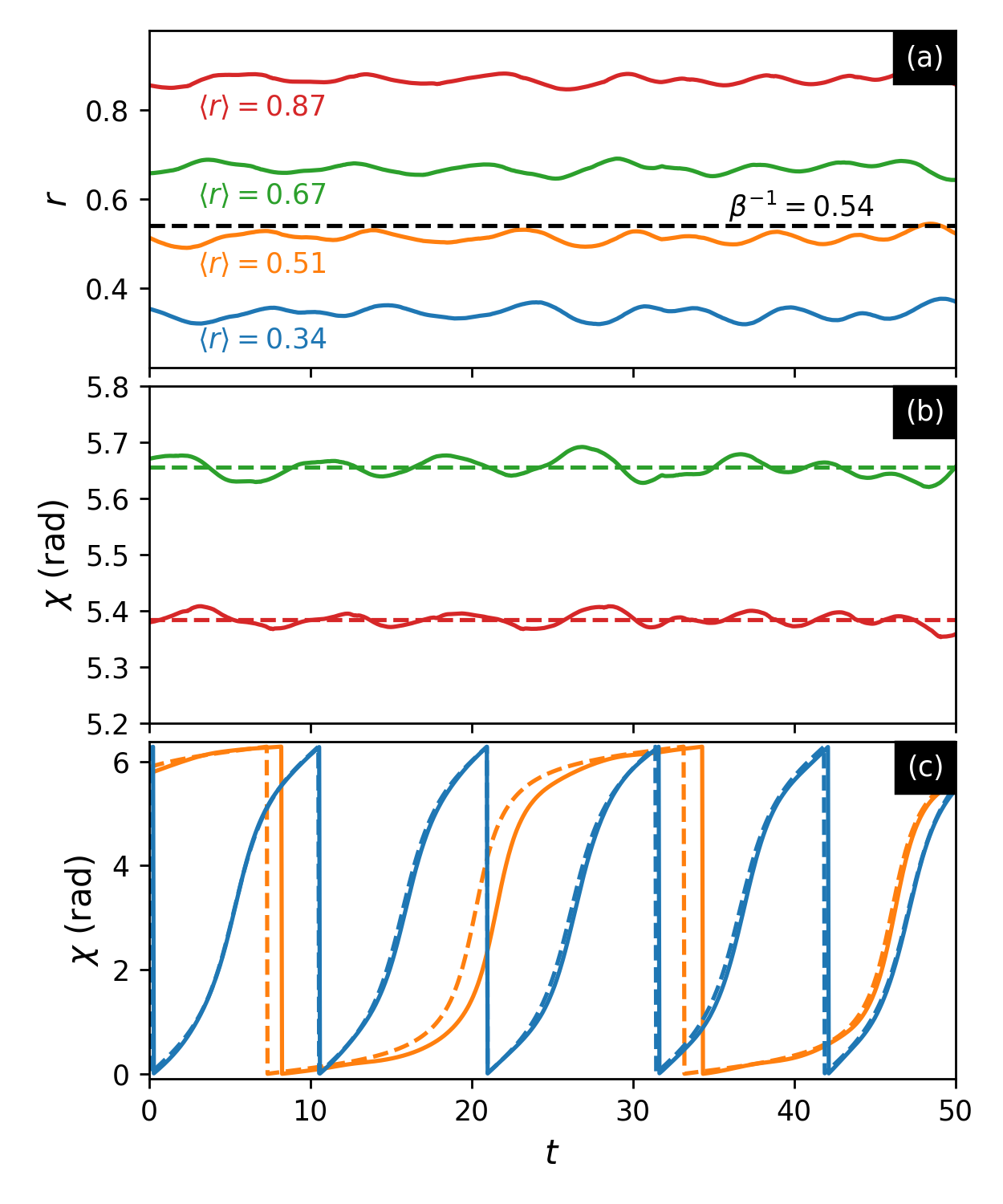}
\caption{Time evolution of (a) radial position and (b, c) tilt angle $\chi=\theta-\phi$ of 4 different particles in a system with $N=500$ and $\beta = 1.85$. The dashed lines in (b) and (c) correspond to the theoretical prediction Eqs.~\eqref{eq:chi_out} and \eqref{eq:chi_in}. 
For particles with mean radial position above $\beta^{-1}$, $\chi$ fluctuates around a constant value as the particle orbits the potential well, 
whereas for particles in the inside region $\chi$ performs full revolutions due to the orientation $\theta$ and radial position $\phi$ no longer having the same frequency.
}
\label{fig:ChiTimeseries}
\end{figure}

To gain deeper insights into the efficacy of the quasi-rigid-body model presented in Sec.~\ref{sec:analytical}, we closely examine the time-dependence of the radial position and the tilt angle $\chi=\theta-\phi$ of four particles, two of them with average radial positions $\av{r}>1/\beta$ and the other two with $\av{r}<1/\beta$, as shown in Fig~\ref{fig:ChiTimeseries}. 
All four particles oscillate radially with similar amplitudes $\sim\sigma/2$. Notably, the oscillations at different mean radial positions are highly correlated, indicating that a collective mode of the cluster is excited. This subtle wobbling of the inner region relative to the outer region, as revealed in Video 4 (ESI), underpins the observed highly correlated radial fluctuations. That is to say, the cluster can no longer be treated as a rigid system. Nevertheless, the behavior of the tilt angle is found to align with the quasi-rigid-body model. Remarkably, despite the uniformity of the angular velocity, it unveils radically different polarization dynamics in both regions: for particles with $\av{r}>1/\beta$, $\chi(t)$ only fluctuates slightly around a constant angle, which is in excellent agreement with the asymptotic limit of Eq.~\eqref{eq:chi_out}; for $\av{r}<1/\beta$, $\chi$ undergoes full rotations periodically, in close agreement with Eq.~\eqref{eq:chi_in}. Deviations from the model can be attributed to the fluctuations in the radial position seen in (a). 

\begin{figure}[t]
%\centering
\includegraphics[width= \linewidth]{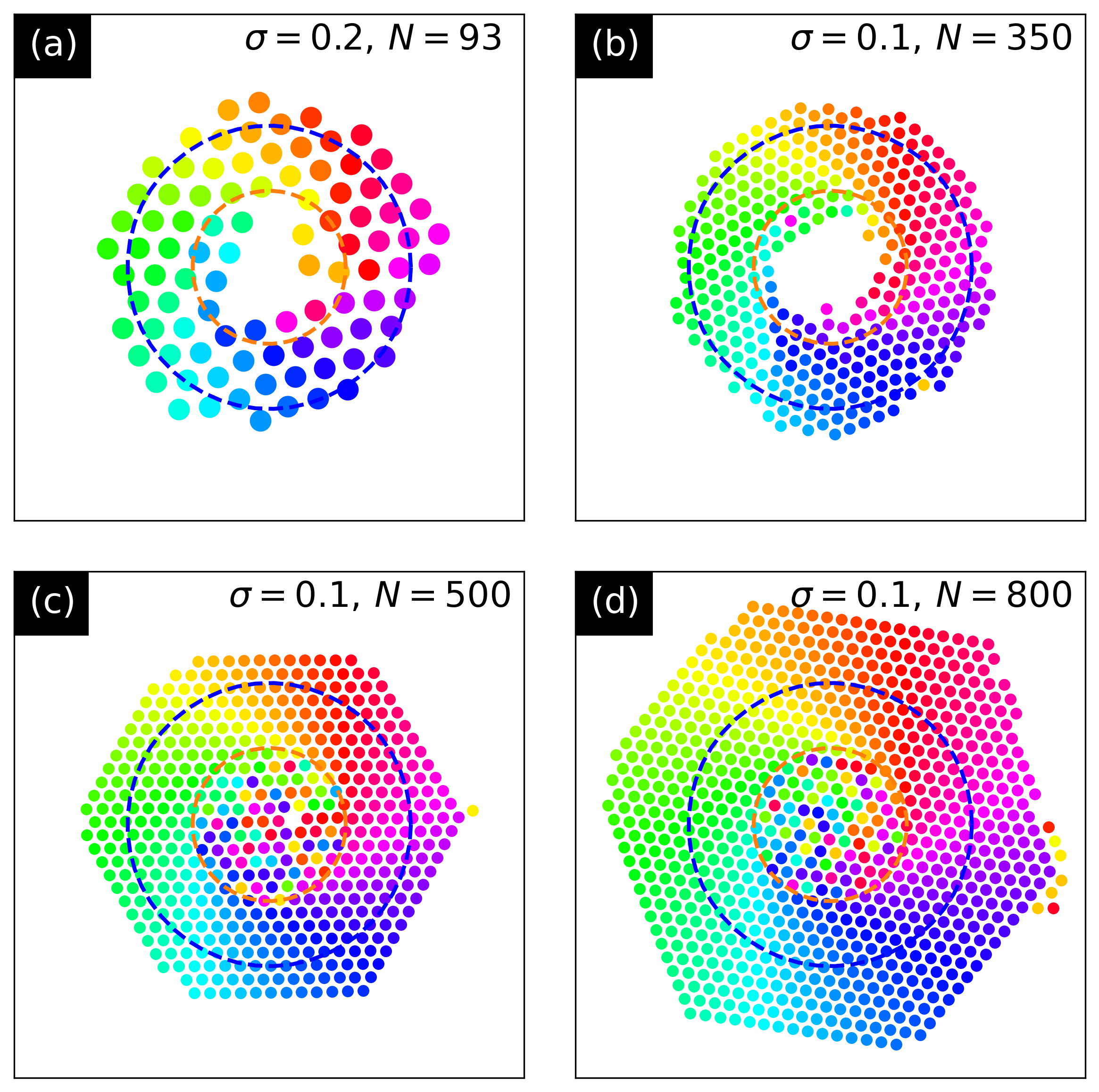}
\caption{ 
Instantaneous configurations for different numbers of particles and $\beta = 1.85$ (see labels in the top right corner of the figures). The dashed blue and orange circles represent the isocline ($r=1$) and the critical radius ($r = 1/\beta$) respectively. The color pattern within the particles depicts their orientation, following the same scheme as described in Fig~\ref{fig:PhasesColor}.}
\label{fig:PhaseExhibition}
\end{figure}

Another striking feature of the UV phase can be observed in Fig.~\ref{fig:PhaseExhibition} (c) and (d), 
where we show snapshots of positions and polar angles $\theta$ of particles for $N=500$ and $N=800$ clusters in the UV phase. In both cases, the particles orientations in the region $r<1/\beta$ are completely random. This can also be understood from Eq.~\eqref{eq:chi_in}. Although particles within the same radial distance have the same oscillation frequency, their phase depends on the value of $\chi$ held by the particle at the moment the system reached its close packed form, $\chi_k(0)$ in Eq.~\eqref{eq:chi_in}. Therefore, the color pattern seen in the region $r<1/\beta$ encodes a memory of how the UV phase was formed in the past. Indeed, the transition between the different phases typically passes through a disordered state, resulting in the observed random phase distribution.  

Altogether, these results suggest that the UV phase rotates not as a strict rigid body but rather as a {\it deformable solid}. Such deformations allow the system to violate the $\dot{\chi}=0$ condition in the region $r<1/\beta$ while still keeping its close-packed structure. Interestingly, a strict rigid-body vortex phase is not possible when particles occupy the $r<1/\beta$ region. Therefore, elastic deformations seem to be essential for the stabilization of the uniform vortex phase.

\subsection{Shear-banding transition}
\label{sec:shear-banding}

\begin{figure}[t]
%\centering
\includegraphics[width= \linewidth]{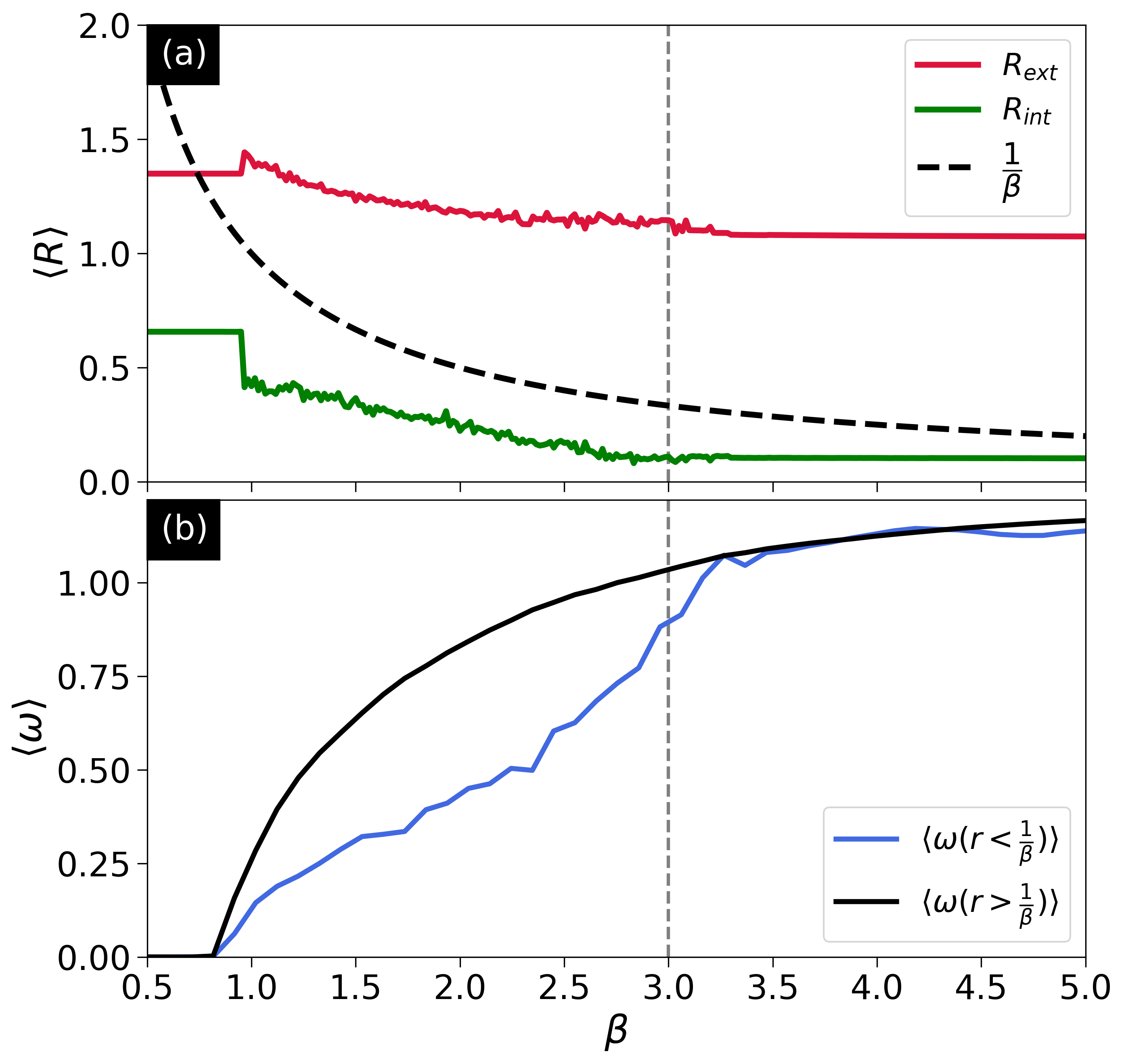}
\caption{SBV to UV transition of a cluster with $N = 93$ and $\sigma = 0.2$. (a) Time-averaged radial position of the outermost ($R_\text{ext}$, red line) and the innermost ($R_\text{int}$, green line) particle in the cluster as a function of $\beta$. The black dashed line marks the critical radius $(1/\beta)$ separating regions of different polar behavior (see text). 
(b) Time-averaged angular speed of the particles belonging to the regions $r>1/\beta$ (black line) and $r<1/\beta$ (blue line). At $\beta = 3$ (grey dashed line), the cluster is nearly fully compacted, and the shear banding phenomenon begins to fade. }
\label{fig:ShearBanding}
\end{figure}

When decreasing beta $\beta$ at fixed $D$ and $f$, the close-packed UV phase becomes unstable with respect to the expansion of the cluster. This can be visualized in Fig.~\ref{fig:ShearBanding} (a), 
where we plot the time-averaged radial positions of the outermost ($R_\text{ext}$) and the innermost ($R_\text{int}$) particles of the cluster as a function of $\beta$ for $D=0$, $N=93$, and $\sigma=0.2$. For $\beta\gtrsim3$, $R_\text{int}<\sigma$ and both $R_\text{int}$ and $R_\text{ext}$ do not change further with $\beta$. This is a result of the cluster being in a fully compact state. However, below this threshold, the cluster expands into a ring with $R_\text{ext}>R_\text{int}>\sigma$. Additionally, the uniformity of the angular speed of the particles, which characterizes the UV phase, breaks down. This is evidenced in Fig.~\ref{fig:ShearBanding}-(b), where we show the time-averaged angular speed of particles belonging to two distinct regions: $r>1/\beta$ and $r<1/\beta$. For $\beta\gtrsim3$, particles in both regions have the same angular speed, as expected for the UV phase. In contrast, for $\beta\lesssim3$, particles occupying the $r<1/\beta$ region have, in average, considerably lower $\omega$, that is, the system is in the SBV phase, with two distinct concentric bands rotating at different angular speeds. 

The concomitant breakdown of cluster compaction and uniformity of the angular speeds, marking the transition from the UV to the SBV phase, can be achieved not only by changing $\beta$, but also by changing the filling fraction.
Fig.~\ref{fig:PhaseExhibition} presents snapshots of clusters with different filling fractions but the same $\beta$ and $D$, (1.85 and 0, respectively). These figures exemplify that, by increasing the filling fraction at a fixed $\beta$, the empty region is progressively populated causing $R_\text{int}$ to decrease to the point where the cluster becomes compact again. At this point, angular velocities become uniform all over the cluster as required for UV rotation.

The time-averaged radial distributions of $P_r$, $P_\phi$, $\dot{\phi}$, and $\dot{\theta}$ for $N=350$ particles and $(\beta,D)=(1.85,0)$ is shown in Fig.~\ref{fig:ByDistance} (c) and (d). The $r>1/\beta$ region is still well described by the quasi-rigid body model, as the particles arrange in an almost crystalline structure in this region. In contrast, the inner band presents highly non-uniform profiles of both, $\dot{\phi}$, and $\dot{\theta}$, thus indicating a fluidlike behavior. Surprisingly, some particles even rotate in the opposite direction of the outer band flow. Another remarkable feature of the inner band is that the particles in this region have in average a considerably high radial polarization, as if they were in a kind of climbing state.

The spontaneous segregation of the system in two rotating bands can be understood as follows. Particles at the inner edge of the ring are pushed away from the cluster towards the center of the potential by the potential itself as well as by the steric forces induced by the other particles. These frustrated particles are constantly trying to fit into the flock. Some of them eventually succeed, thanks to sliding events causing a rearrangement of the system, which involves particles from both bands (see Supplementary Video 4, ESI). In this case, the lucky particle is replaced by another one.
Those that don't succeed are ejected back and the whole process repeats. In order to reach the flock, the stray particles must first climb the confining potential. At this point, they can do that with a slight trend towards either clockwise or counterclockwise rotation. Particles with a clockwise trend have a better chance to get absorbed by the flock if the flock also rotates clockwise and vice-versa, while particles moving opposite to the main flow will remain at the inner band. This process naturally results in an average opposite motion of particles belonging to the inner band, thus explaining the sign change observed in the radial distribution of the angular velocity.

The shear banding transition also occurs when one starts from $\beta<1$ at the climbing phase and increases $\beta$ above a threshold value, $\beta_c\sim1$, at fixed small $D$. Just above threshold, the symmetry is spontaneously broken and the system rotates on average either clockwise or anticlockwise. For moderate filling fractions, $f\sim1$, and just above $\beta_c$, the ring-shaped cluster yields and self-organizes in a shear banded vortex. For large $f$, the radially polarized phase transitions directly to the uniform vortex phase.

\subsection{Multi-cluster phase}
\label{sec:clusters}

We have observed that, for small filling fractions, $f<1$, and moderate values of $\beta$, the system splits into multiple clusters, as illustrated in Fig.~\ref{fig:PhasesColor}-(e).
The resulting configurations and dynamics possess characteristics of both orbital and vortex orders: all particles within a cluster have the same alignment $\bm{n}_i$ and thereby can be treated as a single giant active particle, orbiting around the confinement center following Eqs.~\eqref{eq:polarcluster_R}, \eqref{eq:polarcluster_Omega} and \eqref{eq:polarcluster_theta}; at the same time, the set of several such giant particles can be seen as being in a vortex state, as their polarizations acquire the same azimuthal component, thus leading to the observed coherent rotation of the clusters (see Supplementary Video 5, ESI).
\begin{figure}[t!]
%\centering
\includegraphics[width= \linewidth]{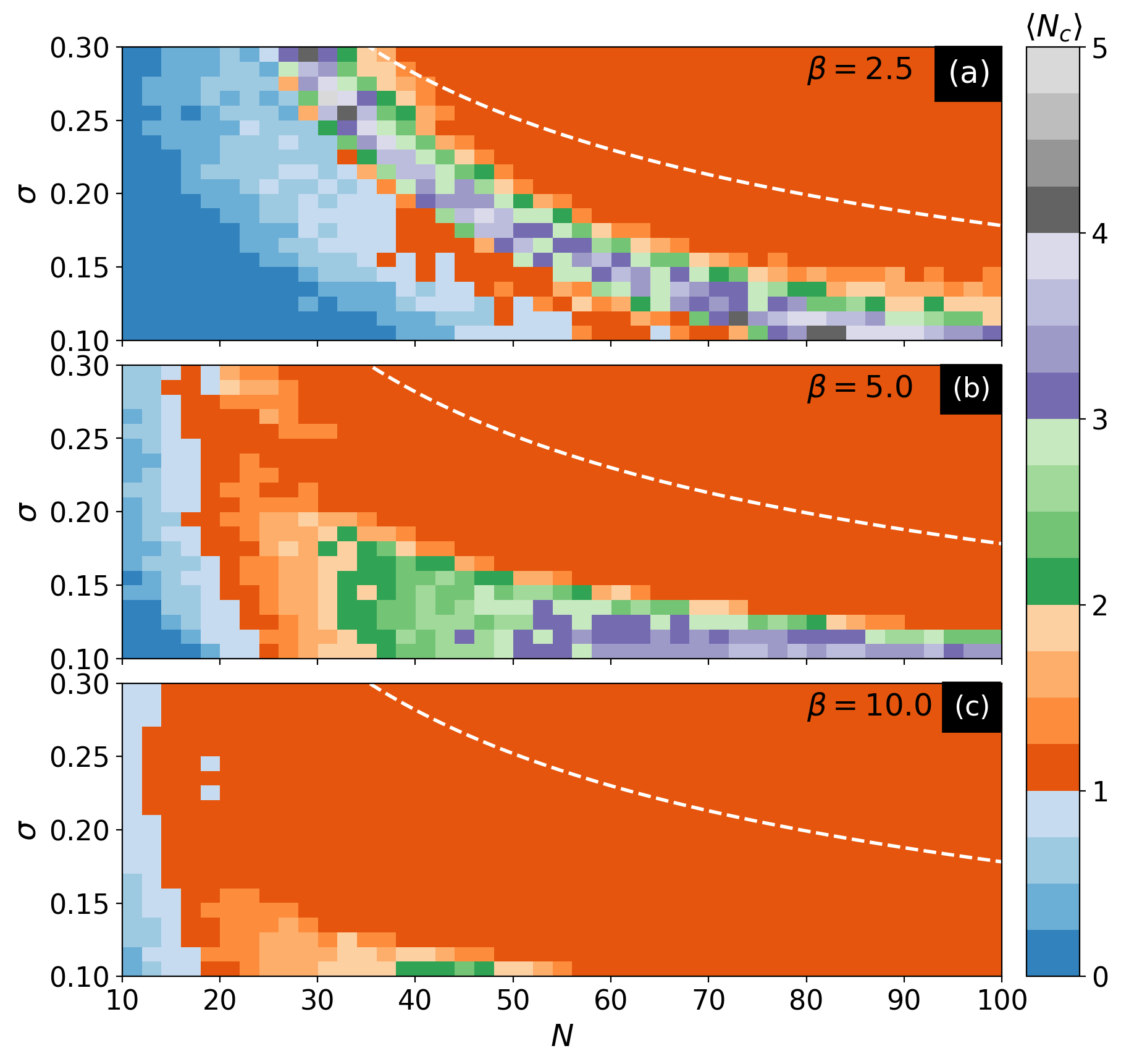}
\caption{Density plots of the average number of clusters as a function of particle size $\sigma$ and total number of particles $N$ for $D=0$ and different values of $\beta$, indicated in upper right corners.
Higher values of $\beta$ lead to tighter systems with only one cluster, even for smaller $N$. 
Dashes represent the $f = 1$ line as defined by Eq~\eqref{eq:FillingFraction}. 
}
\label{fig:ClusterCount}
\end{figure}

To investigate the multi-cluster phase in more detail, we calculate the number of close-packed clusters averaged over 100 independent initial conditions, $\langle N_c\rangle$, exploring the $\sigma N$  parameter space in the range $0.1<\sigma<0.3$ and $10<N<100$. The simulations are performed at $D=0$ for three different values of $\beta$ and the results are depicted in Fig.~\ref{fig:ClusterCount}. Notice that for $f\gtrsim1$, the system typically arrange into a single cluster, while for $f<1$ the particles can either form one or several clusters or none for very disperse systems, depending in an intricate way on $\sigma$, $N$, and $\beta$. At some points, different initial conditions produce different numbers of clusters leading to non-integer $\langle N_c\rangle$.
Notice that higher values of $\beta$ typically induces the formation of a single cluster. For $\beta \gtrsim 10$, the single cluster phase dominates the parameter space. This can be understood as a result of the orbit radius of the clusters, given by $1/\sqrt{\beta}$, becoming so small that different clusters can no longer orbit the potential well without colliding with each other, thereby causing them to merge into a single cluster, which evolves dynamically either in the FM or vortex state. 

\section{Conclusions}
\label{sec:conclusions}

In summary, our analytical and numerical investigations have unveiled a remarkable variety of emergent dynamical phases of harmonically confined self-propelled particles interacting via steric forces and torques. The coupling between torques and the orientational dynamics of the particles, controlled by the angular mobility $\beta$, revealed to be a crucial ingredient for the richness of the phase diagram. For small $\beta$, the particles self-organizes into a static radially polarized state, reminiscent of the climbing phase of single confined self-propelled particles. As one increases $\beta$ past a critical value, the spontaneous symmetry is broken and the system initiates a rotational motion in either the clockwise or counterclockwise direction. The resulting state of the system can be either of several distinct dynamical phases: Orbiting ferromagnetic state, where the cluster orbits the potential center with fully aligned polarizations as in a magnet; Uniform vortex, where the net azimuthal components of the particles orientations add up to rotate the whole system uniformly concentric to the confining center; Shear-banding vortex, where the system segregates in two concentric bands rotating with different angular speeds; and the multicluster phase, where the system divides into several smaller polarized clusters orbiting the potential center. The specific phase exhibited is contingent on several parameters, including filling fraction, noise intensity, and $\beta$, as well as the historical context of parameter changes.

Digging deep into the properties of these phases reveals a few surprising features. For instance, in the uniform vortex phase, the angular speed of the particles is uniformly distributed, but the dynamics of their polarization vectors are highly nonuniform. This might initially appear paradoxical since rigid body rotation typically requires simultaneous evolution of both angular positions and polarization vectors at the same rate. However, our investigation shows that the uniform vortex phase behaves not as a strict rigid body but rather as a deformable solid. This unique characteristic allows for the decoupling of translational and polar dynamics, resolving the apparent contradiction. Moreover, an intriguing observation emerges in the shear-banded vortex phase, where particles in the inner band move in the opposite direction to those in the outer band. This counterflow results from a filtering effect, where particles from the inner band are more likely to be absorbed by the outer band if they move in the same direction, leading to an overpopulation of particles rotating in the opposite direction in the inner band region.

An intriguing avenue for further research on this system involves a deeper investigation into the deformable properties of the vortex phase, specially considering the relevance of deformable active systems for smart soft materials and their potential application in understanding biological functions~\cite{Manning2023}. Specifically, focusing on identifying possible selection mechanisms of collective modes~\cite{baconnier2022,baconnier2023} and exploring whether external excitation of such modes could induce a transition from a vortex to the ferromagnetic state could unveil novel insights. Another crucial aspect deserving dedicated attention is the role of defects within the observed vortex states. In our simulations, these defects in the polarization space appear stable over the entire simulated time window, raising questions about their translocation and potential annihilation. A detailed study addressing how such defects could be manipulated or eliminated would contribute significantly to our understanding of the stability and dynamics of vortex states in confined active matter systems.

\begin{acknowledgments}

This work was financed in part by Coordenação de Aperfeiçoamento de Pessoal de Nível Superior - Brasil (CAPES), Finance Code 001. CCSS is funded by Conselho Nacional de Desenvolvimento Científico e Tecnológico - Brasil (CNPq), Grant No. 312240/2021-0.

\end{acknowledgments}

\bibliography{Activematter}% Produces the bibliography via BibTeX.

\end{document}

\title{Polar order, shear banding, and clustering in confined active matter: Supplementary Information}
%\homepage{\label{ESI}Electronic Supplementary Information (ESI) available.}

\author{Daniel Canavello, Rubens H. Damascena, Leonardo R. E. Cabral, \\and Clécio C. de Souza Silva}
%\affiliation{Departamento de Física, Centro de Ciências Exatas e da Natureza, Universidade Federal de Pernambuco, Recife--PE, 50670-901, Brasil}
\date{\today}

% \begin{abstract}
%     This supplementary information contains  
% \end{abstract}
\maketitle
\tableofcontents

\section{Calculation details of the deterministic steady-state solutions of
close-packed clusters section} %{Collective motion of active particles clusters}
\label{SI.sec:model}
Here we present in more detail the analytical results regarding the motion of the system in the absence of noise. We start with the equations of motion of the system:
%Eqs.~\eqref{eq:motion} and~\eqref{eq:orientation-part} are given by:
\begin{align}
  & \dot{\bm{r}}_k +  \bm{r}_k - \sum_{\substack{l=1\\l \neq k}}^N {\bm{f}_{kl}} = \bm{n}_k,
  \label{eq.1A}\\
  & \dot{\theta}_k = \beta (\bm{n}_k\times \dot{\bm{r}_k})\cdot \hat z 
  = -\beta \CC{\bm{n}_k\times \PP{\bm{r}_k - \sum_{\substack{l=1\\l \neq k}}^N {\bm{f}_{kl}}}}\cdot \hat z,\label{eq.1B}
\end{align}
for parabolic confinement and arbitrary pairwise inter-particle forces given by:
\begin{eqnarray}
\bm{f}_{kl} = f(r_{kl}) \dfrac{\bm{r}_{kl}}{r_{kl}},\qquad \text{ for }  k,\,l = 1,\,2,\,\ldots N ,
\label{eqSI:inter_forces}
\end{eqnarray}
are the particles indexes, $\bm{r}_k = \PP{r_k \cos\phi_k,\,r_k \sin\phi_k}$, $\bm{n}_k = \PP{\cos\theta_k,\,\sin\theta_k}$, $\bm{r}_{kl} = \bm{r}_k - \bm{r}_l$, and $r_{kl}  = |\bm{r}_k - \bm{r}_l|$.

Now, let us investigate a general rigid body motion of the system. In this case, it is well known that the system motion consists of a translation plus a rotation of the entire system. Therefore, the position of the $k^{\rm th}$ particle of the system can be described by 
\begin{align}
\bm{r}_k = \bm{R} + \Mrs{S}\PP{\phi}\bm{s}_k,
\label{eqAp:2}
\end{align}
where $\bm{R} = \sum_{k=1}^N \bm{r}_k / N$ is the position of the system centroid, %$\Mrs{S}\PP{\phi}$ is the rotation matrix, 
\begin{align}
\Mrs{S}\PP{\phi} = \PP{
\begin{array}{cc}
\cos\phi & -\sin\phi\\
\sin\phi & \cos\phi
\end{array}
},
\label{eqAp:rotmatrix}
\end{align}
is the rotation matrix,
$\phi$ is a rotation angle in a frame whose origin is at the centroid position, and $\bm{s}_k$ is the particle position in the moving frame. Substituting this equation in Eq.~\eqref{eq.1A} we obtain
\begin{align}
  & \dot{\bm{R}} + \bm{R} + \Mrs{S}(\phi) \CC{
  \dot{\bm{s}}_k + \omega \hat{z} \times \bm{s}_k  
  - \sum_{\substack{l=1\\l \neq k}}^N \dfrac{f(s_{kl})}{s_{kl}} \bm{s}_{kl}
  } = \bm{n}_k,
%  \dot{\bm{r}}_k +  \bm{r}_k - \sum_{\substack{l=1\\l \neq k}}^N {\bm{f}_{kl}} = \bm{n}_k
  \label{eqAp.1C}
\end{align}
where $\omega = \der{\phi}{t}$, $\bm{s}_{kl} = \bm{s}_{k}-\bm{s}_{l}$, and ${s}_{kl} = |\bm{s}_{k}-\bm{s}_{l}|.$ By summing the contributions of all particles to Eq.~\eqref{eqAp.1C}, the internal forces cancel out and we find
\begin{align}
  & \dot{\bm{R}} + \bm{R} = \dfrac{1}{N}\sum_{k=1}^N \bm{n}_k = \av{\bm{n}}.
  \label{eqAp.1D}
\end{align}
As a consequence the centroid motion is governed by the average of the orientation forces of the system. Substituting back this equation in Eqs.~\eqref{eqAp.1C} and ~\eqref{eq.1B} results in
\begin{align}
  & \dot{\bm{s}}_k + \omega \hat{z} \times \bm{s}_k  
  - \sum_{\substack{l=1\\l \neq k}}^N \dfrac{f(s_{kl})}{s_{kl}} \bm{s}_{kl}
   = \Mrs{S}(-\phi) \PP{\bm{n}_k - \av{\bm{n}} },
  \label{eqAp.1E}\\
  & \dot{\theta}_k = \beta\CC{ \PP{\bm{n}_k \times \dot{\bm{R}}}\cdot \hat z 
+ \PP{\boldsymbol{\eta}_k\times \dot{\bm{s}}_k}\cdot \hat z + \omega \PP{\boldsymbol{\eta}_k\cdot \bm{s}_k} }. 
\label{eqAp.1F}
\end{align}
For $\dot{\bm{s}}_k = 0$ in the moving frame, the first equation is time independent if either (i) $\bm{n}_k = \av{\bm{n}}$ for all particles %(i.e., all the particles orientations are equal to $\av{\bm{n}}$) 
or (ii) $\boldsymbol{\eta}_k = \Mrs{S}(-\phi)\bm{n}_k$ is time independent (all orientations rotate at the same angular velocity $\omega$ together with the whole configuration). In the former case, all the particles have the same orientation angle, that is  $\theta_i = \theta_{\rm c}$ for $i =1 ,\,2,\,\ldots N$, while in the latter the time derivative of the particles orientations are the same, i.e., $\dot{\theta}_k = \dot{\theta}_l = \omega$ for any pair of particles in the system rotating as a rigid body with angular velocity $\omega$. Hence, any of these conditions are sufficient to warrant a rigid body motion (with no deformations) of the system. 

The two particular cases above result in Eq.~\eqref{eqAp.1F} to be rewritten as: 
\begin{itemize}
\item 
For $\bm{n}_k = \av{\bm{n}} = \bm{n}_{\rm c}$ (which is equivalent to $\theta_i = \theta_{\rm c}$)  yields 
\begin{eqnarray}
\PP{\bm{n}_{\rm c} \times \dot{\bm{R}}}\cdot \hat z 
 + \omega \boldsymbol{\eta}_{\rm c}\cdot \bm{s}_k = \beta^{-1}\dot{\theta}_{\rm c};
 \label{eqAp.polarized0}
\end{eqnarray}
\item For $\dot{\theta}_k = \dot{\theta}_l = \omega$,
we have:
\begin{eqnarray}
\PP{\bm{n}_{k} \times \dot{\bm{R}}}\cdot \hat z 
 + \omega \PP{\boldsymbol{\eta}_{k}\cdot \bm{s}_k - \beta^{-1}} = 0.
 \label{eqAp.vortex0}
\end{eqnarray}

\end{itemize}
Therefore, for those conditions above if both the centroid motion and the rigid body rotation are present, they are intertwined in a way given by either Eq~\eqref{eqAp.polarized0} or Eq.~\eqref{eqAp.vortex0}. 

In the following sections we analyse two particular cases of interest related to the above conditions associated with rigid body motion.

\subsection{Polarized phase}
\label{SI.ssec:polar}

We denote the case $\bm{n}_k = \av{\bm{n}} = \bm{n}_{\rm c}$ (i.e., $\theta_k = \theta_{\rm c}$  for all $k$) for all particles the \emph{polarized state}. Let us consider one particular case in which $\omega = 0$ (i.e., the cluster does not rotate around the centroid position, $\bm{R}$). In this case, the system behaves as `one particle' and Eq.\eqref{eqAp.polarized0} becomes 
\begin{align}
-\dot{R}\sin\chi_{\rm c} + R \dot{\varphi} \cos\chi_{\rm c} & = \beta^{-1}\dot{\theta}_{\rm c},
\label{eqAp.orientPol0}
\end{align} 
where $\chi_{\rm c} = \theta_{\rm c} - \varphi$ is the tilt angle and $\varphi$ is the angle $\bm{R}$ makes with the horizontal axis. Meanwhile, Eq.~\eqref{eqAp.1D} projected along and orthogonally to $\bm{R}$ gives,
\begin{align}
\dot{R}+R & = \cos\chi_{\rm c},\label{eqAp.radialPol} \\
%\PP{\bm{R} \times \dot{\bm{R}}} \cdot \hat z  = R^2 \Omega & =  R \sin \chi_{\rm c}.
R \, \dot{\varphi} & =  \sin \chi_{\rm c}.
\label{eqAp.orthoPol}
\end{align} 
Substituting these equations in Eq.~\eqref{eqAp.orientPol0}, we obtain
\begin{align}
R^2  \dot{\varphi} & = \beta^{-1}\dot{\theta}_{\rm c}.
\label{eqAp.orientPol0p5}
\end{align} 
This equation also shows that the areolar speed covered by the centroid is equal to the time derivative of the its orientation, $\theta_{\rm c}$, divided by $\beta$. This is a particular case of a more general result for active particles in elliptic harmonic confinement.\cite{Damascena22}

The above equations are the same as the ones for one active particle in a harmonic circular confinement. Therefore, for $\beta > 1$ a time independent periodic solution is given by $\dot{R} = 0$ and $\dot{\varphi} = \dot{\theta}_{\rm c}$, that is the centroid performs a periodic circular motion with angular velocity, $\Omega = \der{\varphi}{t}$. For such a case, we have
\begin{align}
R & = \beta^{-1/2}, \label{eqAp.pol_R}\\
\Omega & = \pm \sqrt{\beta - 1}, \label{eqAp.pol_Omega}\\ 
\chi_{\rm c} & = \theta_{\rm c} - \varphi = \arccos\PP{\beta^{-1/2}} \quad \Rightarrow\quad \theta_{\rm c} = \arccos\PP{\beta^{-1/2}} \pm\sqrt{\beta - 1} t  + \theta_{\rm c}\PP{0}.\label{eqAp.pol_theta}
\end{align} 
%which are the same solutions for one active particle in a harmonic circular confinement,\cite{Damascena22,DauchotPRL19} as one might already have expected.

\subsection{Vortex phase}
\label{SI.ssec:vortex}

Consider the case in which $\dot{\theta}_k = \omega$ for all the particles and assume that the particles rotate around the origin (that is $\bm{R} = 0$ and $\bm{r}_k = \bm{s}_k$). In this case, Eq.~\eqref{eqAp.1D} tell us that $\av{\bm{n}} = 0$. Now, let us compute the angular velocity for this situation of rigid body motion. By taking the vector product between $\bm{r}_k$ and Eq.~\eqref{eq.1A} and summing over the contributions of all the partilces, internal torques cancel out we find
\begin{align}
\sum_k s_k^2 \dot{\phi}_k = \omega \sum_k s_k^2 = \sum_k s_k \sin{\chi}_k,
\label{eq.Ap_omegaRB0}
\end{align} 
where $\chi_k = \theta_k - \phi_k.$
From Eq.~\eqref{eqAp.vortex0} we have $\boldsymbol{\eta}_k\cdot \bm{s}_k = s_k \cos\chi_k = 1/\beta$. Therefore, 
\begin{align}
\omega  = \dfrac{\displaystyle\sum_k \sqrt{s_k^2 - \dfrac{1}{\beta^2}} }{\displaystyle\sum_k s_k^2}.
\label{eq.Ap_omegaRB1}
\end{align} 
This result shows us that a strict rigid body motion is possible in the vortex phase only for particles at $r > 1/\beta$.

Now let us relax the condition $\dot{\theta}_k = \omega$, which means the system is not in strict rigid body motion, while still assuming that the particles rotate around the origin  with approximate fixed radial positions (i.e., $\bm{r}_k = \bm{s}_k$ with $\dot{s}_k\approx  0$) and angular velocity $\dot{\phi}_k \approx \omega$. Thus, Eq.~\eqref{eqAp.1F} becomes,
\begin{align}
\dot{\chi}_k = \PP{\beta s_k \cos\chi_k - 1}\omega,
\label{eqAp.Vortex1}
\end{align}
whose solution is given by,
~\cite{AbraSteg}
\begin{align}
-\omega \,t & = \int_{\chi_{k}(0)}^{\chi_{k}(t)}
\dfrac{d\chi}{1 - \beta s_k \cos\chi} = 
 \dfrac{\CC{G(t) - G(0)}}{\sqrt{\bb{\beta^2 s_k ^2 - 1}}}
\label{eqAp.chiVortex1}
\end{align}
where $\chi_{k}(0)$ is the value of $\chi_{k}$ at some initial time $t = 0$ and,
\begin{align}
G(t) = 
\left\{
\begin{array}{ l l}
2\, \arctan\CC{
\dfrac{\PP{1+\beta s_k} \tan(\chi/2)}{\sqrt{\bb{\beta^2 s_k^2 - 1}}}
}, & \text{ for } \beta s_k < 1, 
\\
%-\cot\PP{\chi/2}, & \text{ for } \beta R = 1,\\
\ln\CC{
\dfrac{
\sqrt{\bb{\beta^2 s_k^2 - 1}} - \PP{1 + \beta s_k}\tan(\chi/2)
	}{
\sqrt{\bb{\beta^2 s_k^2 - 1}} + \PP{1 + \beta s_k}\tan(\chi/2)	
	}
} , & \text{ for } \beta s_k > 1.
\end{array}
\right.
\label{eqAp.chiPol2}
\end{align}
The above solution gives rise to two distinct time dependences of $\chi_k$. 
For $s_k > 1/\beta$ there is a monotonic dependence of $\chi_k$ on time, 
\begin{align}
\chi_k = 2\arctan\lC\lP 
 \dfrac{C_+ + C_- e^{-t/\tau_\omega}}{C_+ - C_- e^{-t/\tau_\omega}}\rP
\sqrt{\dfrac{\beta s_k-1}{\beta s_k+1}}\rC, \label{eq.A8BSI}
\end{align}
where \(C_\pm = \tan\lP\chi_k(0)/2\rP \pm \sqrt{\lP \beta s_k-1\rP/\lP \beta s_k+1\rP}\) and 
\(\tau_\omega = 1/\omega\sqrt{\beta^2 s_k^2 - 1}\). In this case, for $t \gg \tau_\omega$ we have the asymptotic value \(\chi_k = 2\arctan\sqrt{\lP \beta s_k-1\rP/\lP \beta s_k+1\rP}\), which gives  \(\cos \chi_k = 1/\beta s_k\). This is exactly the result expected by assuming $\omega = \dot{\theta}_k$.

On the other hand, $\chi_k$ has oscillatory response for $r_k < 1/\beta$, given by 
\begin{align}
\chi_k = -2\arctan\lC 
\sqrt{\dfrac{1-\beta s_k}{1+\beta s_k}} \tan\lP
\sqrt{1-\beta^2 s_k^2}\dfrac{\omega t}{2} - C_0
\rP
\rC, \label{eq.A8ASI}
\end{align}
where $C_0 = \arctan\lC 
\sqrt{\lP 1-\beta s_k\rP/\lP 1+\beta s_k\rP} \tan\lP\chi_k(0)/2\rP\rC$. % and $\chi_k(0)$ the initial value of $\chi_k$ at $t = 0$.

These above expressions for $\chi_k$ allow us to obtain the time averages of $\bm{n}_k\cdot \bm{r}_k = s_k \cos\chi$ and $\PP{\bm{r}_k\times \bm{n}_k}\cdot \hat z = s_k \sin\chi$. For that we compute the asymptotic limit of $\cos\chi_k$ and $\sin\chi_k$ for $r_k = s_k > 1/\beta$ and their time averages for $r_k = s_k < 1/\beta$. As already mentioned, for $s_k > 1/\beta$ and $t \gg 1/\omega\sqrt{\beta^2 s_k^2 - 1}$, we have \(\cos \chi_k = 1/\beta s_k\), which gives
\begin{align}
\bm{n}_k\cdot \bm{r}_k = s_k \cos\chi_k = \dfrac{1}{\beta} \qquad \text{ and } \qquad
\PP{\bm{r}_k\times \bm{n}_k}\cdot \hat z =  s_k \sin\chi_k = \sqrt{s_k^2 - \dfrac{1}{\beta^2}}.
\label{eq.Ap_Pr_Pphi_outer}
\end{align}
%(Remind that in the vortex phase we assumed $r_k = s_k$ since the centroid is at the origin.)
For $s_k < 1/\beta$, after some straightforward calculation Eq.~\eqref{eq.A8ASI} gives
\begin{align}
\sin\chi_k & =  -\dfrac{
\sqrt{1-\beta^2s_k^2}\sin\PP{
\sqrt{1-\beta^2s_k^2}\omega t - 2C_0
}
}{
1+\beta s_k\cos\PP{
\sqrt{1-\beta^2s_k^2}\omega t - 2C_0
}
},\label{eq.Ap_sinchi_in}\\
\cos\chi_k & = \dfrac{
\beta s_k+\cos\PP{
\sqrt{1-\beta^2s_k^2}\omega t - 2C_0
}
}
{
1+\beta s_k\cos\PP{
\sqrt{1-\beta^2s_k^2}\omega t - 2C_0
}
}.  \label{eq.Ap_coschi_in}
\end{align}
The time average of $\sin\chi_k$ over one period of time $T = 2\pi/\omega\sqrt{1-\beta^2 s_k^2}$ is zero since it is an odd function. On the other hand, the time average of $\cos\chi_k$  is nonzero and can be computed by using standard table of integrals~\cite{gradshteyn}. Therefore,
\begin{align}
\av{\sin\chi_k}_T = 0\qquad \text{ and } \qquad
\av{\cos\chi_k}_T = \dfrac{
1 - \sqrt{1 - \beta^2 s_k^2}
}{
\beta s_k}.
\label{eq.Ap_trig_in_average}
\end{align}
Consequently, for $s_k < 1/\beta$,
\begin{align}
\av{\bm{n}_k\cdot \bm{r}_k}_T = \dfrac{
1 - \sqrt{1 - \beta^2 s_k^2}
}{
\beta} \qquad \text{ and } \qquad
\av{\PP{\bm{r}_k\times \bm{n}_k}\cdot \hat z}_T = 0.
\label{eq.Ap_Pr_Pphi_inner}
\end{align}
Now, let us calculate the average angular velocity taking into account the above results. Again, by taking the vector product between $\bm{r}_k$ and Eq.~\eqref{eq.1A}, summing over the contributions of all the particles, and canceling out the internal torques, we find
\begin{align}
\sum_k s_k^2 \dot{\phi}_k = \sum_k s_k \sin{\chi}_k.
\label{eq.Ap_omega0}
\end{align} 
Here, we use $\dot{\phi}_k \approx \omega$ and take the time average of this equation. Since $\av{\PP{\bm{r}_k\times \bm{n}_k}\cdot \hat z}_T = s_k \av{\sin\chi_k}_T = 0$ for $r_k = s_k < 1/\beta$, Eq.~\eqref{eq.Ap_omega0} results in
\begin{align}
\av{\omega}_T = \dfrac{\displaystyle\sum_k{}^{\prime}\, s_k \av{\sin{\chi}_k}_T}{\displaystyle\sum_k s_k^2 } = \dfrac{\displaystyle\sum_k{}^{\prime}\, \sqrt{
s_k^2 - \dfrac{1}{\beta^2}
}}{\displaystyle\sum_k s_k^2 },
\label{eq.Ap_omega1}
\end{align}
where $\displaystyle\sum_k{}^{\prime}$ means sum over only the particles outside of $r = 1/\beta$, while $\displaystyle\sum_k$ is the sum over all the particles. This expression is similar (but not equal) to Eq.~\eqref{eq.Ap_omegaRB1}.

Finally, we return to Eq.~\eqref{eqAp.Vortex1} and take its time average. It furnishes the time average of the particles orientation angle,
\begin{align}
\av{\dot{\theta}_k}_T = \beta s_k \av{\cos\chi_k}_T\av{\omega}_T =  \lCB
\begin{array}{rr}
\av{\omega}_T, & \text{ for } s_k > \beta^{-1}\\
\CC{1 - \sqrt{1-\beta^2s_k^2}}\av{\omega}_T, & \text{ for } s_k < \beta^{-1}
\end{array}
\rp
\label{eq:SI_dot-theta}
\end{align}

These obtained time dependencies of $\chi_k$ show that, although for $s_k > 1/\beta$ the conditions for a strict rigid body motion can be realized, that is not the case for particles inside $s < 1/\beta$. This conclusion can be seen by noticing that a time dependent $\chi_k$ implies in a time dependence of $\bm{n}_k$, which also implies that both $\dot{r}_k$ and $\dot{\phi}_k$ in Eq.~\eqref{eq.1A} do depend on time. Therefore, the above results for the vortex phase are to be taken as an approximate model for the dynamics of particles moving together with the same angular velocities and with small variations of their radial distances to the origin.

\newpage
\section{Supplementary Figures}
\subsection{Effect of translational noise $D_t$}
\begin{figure}[htb!]
\centering
\includegraphics[width= 0.5\linewidth]{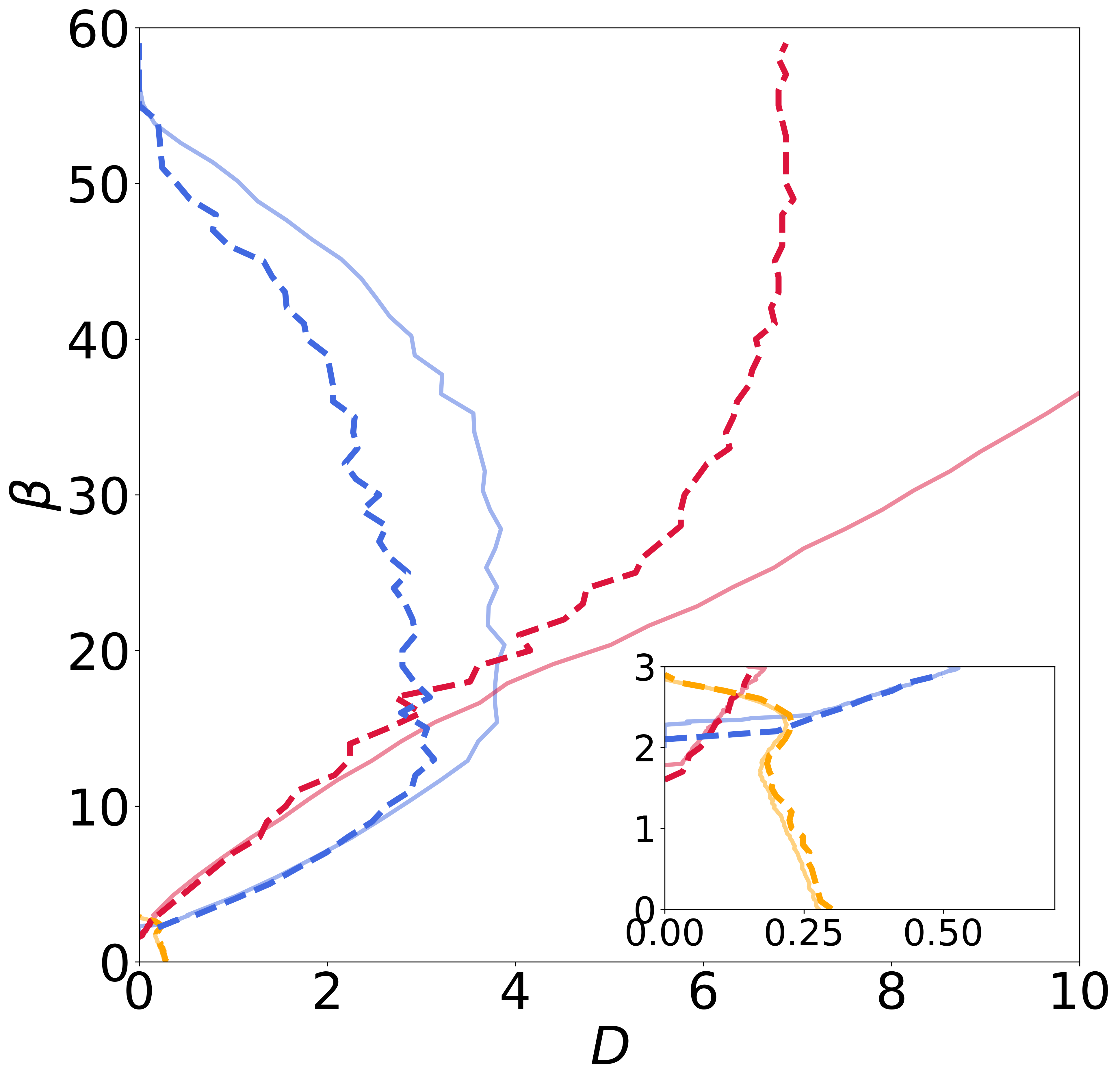}
\caption{0.5 contour levels of the averaged total (red), azimuthal (blue), and radial (orange) order parameters. The solid lines represent the $D_t=0$ case considered in the main text, that is, the same as in Fig. 2. The dashed lines represent the case where translational noise $D_t=\sigma^2D/3$ is added.}
\label{fig:DTrans}
\end{figure}

In this analysis, we explore a scenario where thermal noise, rather than the self-propulsion mechanism, serves as the primary source of fluctuations. 
The translational ($D_t$) and rotational ($D$) diffusion constants in this case are related by $D_t = \sigma^2D/3$. Since for most cases of interest (our work included) particle size is small compared to other relevant length scales, $D_t$ is typically small compared to $D$. 
On the other hand, the proportionality between both diffusion constants in the thermal noise case warrants that increasing $D$ increases the importance of the translational noise. 
As illustrated in the figure, translational noise notably influences phase boundaries, particularly for larger $D$, with more drastic changes observed in the high $\beta$ region of the phase diagram. 
Conversely, for $D \lesssim 1$, the phase boundaries remain essentially unchanged, whether or not $D_t$ is present. 
This indicates that translational noise has a negligible effect on the radially polarized and shear-banded vortex phases. 
Despite alterations in phase boundaries, all phases maintain the same qualitative features observed in the case of active noise, and no new phases were detected.

\newpage
\section{Supplementary Videos}

In all videos the dashed blue lines indicate the critical isocline, while the orange dashed lines in videos 3 and 4 indicate the $1/\beta$ line. %Below, the parameter values for each video are denoted as $(\beta, D, N, \sigma)$. 

\subsection*{Video 1: Radially Polarized (RP) phase (\tt{Video1-RP.mp4})}
This video captures the dynamic behavior of the radially polarized (RP) phase when a small noise ($D=0.05$) is added. Other parameters are $\beta=0.5$, $N=350$, and $\sigma=0.1$. The right side displays real-time polar order parameters of the system.  The added noise results in a highly fluid behavior, while still retaining a high degree of radial polar order, as corroborated by the high values of $P_r$.

\subsection*{Video 2: Orbiting Ferromagnetic (FM) phase (\tt{Video2-FM.mp4})}

This video shows the evolution of the FM phase for $(\beta, D, N, \sigma)=(10.0, 1.0, 350, 0.1)$. Despite the considerably high noise, the phase remains qualitatively unchanged with respect to the noiseless case, showcasing its superior noise tolerance compared to other ordered phases.

\subsection*{Video 3: Uniform vortex (UV) phase (\tt{Video3-UV.mp4})}
This video presents two distinct perspectives of the UV phase captured in a single simulation run with parameters $(\beta, D, N, \sigma)=(1.85, 0.0, 500, 0.1)$: (left) laboratory-frame view, with colors representing the orientation angles ($\theta$) of the particles; and (right) rotating-frame view, with colors representing  the tilt angles ($\chi=\theta-\phi$) relative to the radial axis. The video highlights the markedly different behavior of polar orientations for particles within regions $r>1/\beta$ compared to those within $r<1/\beta$. Our quasi-rigid-body model accurately captures this distinction, as discussed in the main text. The rotating-frame view provides a clearer visualization of collective deformations within the cluster, revealing a subtle wobbling of the inner region relative to the outer region.  

\subsection*{Video 4: Shear-banded vortex (SBV) phase (\tt{Video4-SBV.mp4})}
This video illustrates both lab-frame and rotating-frame views of the Shear-banded vortex (SBV) phase in a simulation run with $(\beta, D, N, \sigma)=(1.85, 0.0, 350, 0.1)$. Notably, particles in the inner region rotate on average in the opposite direction to the outer bulk, defining the characteristic emergent shear banding. The rotating frame view enhances the visualization of crucial points discussed in the main text: (i) The reddish shades of particles in the inner region (that is, $\chi\simeq0$) reveal their tendency to align radially; (ii) Particles in the bulk (outer region) undergo sporadic rearrangements, facilitating the exchange of particles between both regions. These features are essential for understanding the shear-banding transition.

\subsection*{Video 5: Multi-cluster phase (\tt{Video5-Clusters.mp4})}
This video illustrates the evolution of the multi-cluster phase for $(\beta, D, N, \sigma)=(5.0, 0.0, 93, 0.1)$. Here, the system separates into 4 different clusters, each one showcasing traits similar to the FM phase, but orbiting the potential well as a vortex phase. The global vortex-like behavior is evidenced by the high values of $P_\phi$. 
Note that $|\bm{P}|$ is close to zero since we are taking all particles into account, and thus the clusters in opposing sides cancel each other. The value of $|\bm{P}|$ considering only the particles inside a cluster is close to $1$.